%% file: paper.tex
\PassOptionsToPackage{dvipsnames}{xcolor}
\PassOptionsToPackage{table}{xcolor}
\documentclass[sigconf,nonacm]{acmart}

\input{cmd/packages.tex}
\input{cmd/commands.tex}

\input{cmd/definitions.tex}

\setcopyright{none}

\begin{document}

	\title{Bytecode-centric Detection of Known-to-be-vulnerable Dependencies in Java Projects}

	\author{Stefan Schott}
	\affiliation{%
		\institution{Paderborn University}
		\city{Paderborn}
		\country{Germany}
	}
	\email{stefan.schott@upb.de}
	\orcid{0000-0002-0644-3297}
	
	\author{Serena Elisa Ponta}
	\affiliation{%
		\institution{SAP Labs}
		\city{Mougins}
		\country{France}
	}
	\email{serena.ponta@sap.com}
	\orcid{0000-0002-6208-4743}
	
	\author{Wolfram Fischer}
	\affiliation{%
		\institution{SAP SE}
		\city{Stuttgart}
		\country{Germany}
	}
	\email{wolfram.fischer@sap.com}
	\orcid{0000-0001-8127-8837}
	
	\author{Jonas Klauke}
	\affiliation{%
		\institution{Paderborn University}
		\city{Paderborn}
		\country{Germany}
	}
	\email{jonas.klauke@upb.de}
	\orcid{0000-0001-9160-9636}
	
	\author{Eric Bodden}
	\affiliation{%
		\institution{Paderborn University \& \\ Fraunhofer IEM}
		\city{Paderborn}
		\country{Germany}
	}
	\email{eric.bodden@upb.de}
	\orcid{0000-0003-3470-3647}

	\begin{abstract}
		\input{sections/abstract.tex}
	\end{abstract}
	
	
	
	\maketitle

	\input{sections/introduction.tex}
	\input{sections/background.tex}
	\input{sections/concept.tex}

	\input{sections/evaluation.tex}

	\input{sections/threats.tex}
	\input{sections/relatedWork.tex}
	\input{sections/conclusion.tex}
	
	\section*{Acknowledgement}
	This work was partially supported by the German Research Foundation (DFG) within the Collaborative Research Centre ''On-The-Fly Computing`` (GZ: SFB 901/3) under the project number 160364472.

	\bibliographystyle{plainurl}
	\bibliography{literature}
\end{document}

%% file: cmd/packages.tex
\usepackage{xspace}
\usepackage{listings}
\usepackage{subcaption}
\usepackage{hyperref}
\PassOptionsToPackage{hyphens}{url}
\usepackage{url}
\usepackage{framed}
\usepackage{xcolor}
\usepackage{amsmath}
\usepackage{multirow}
\usepackage{pifont}
\usepackage{enumitem}

%% file: cmd/commands.tex
\newcommand{\jaralyzer}{\textsc{Jaralyzer}\xspace}
\newcommand{\steady}{\textsc{Eclipse Steady}\xspace}
\newcommand{\jess}{\textsc{Jess}\xspace}

\newcommand{\kb}{\textsc{Project KB}\xspace}
\newcommand{\owaspDC}{OWASP DC\xspace}
\newcommand{\achilles}{\textsc{Achilles}\xspace}

\newcommand{\cmark}{\ding{51}}%
\newcommand{\xmark}{\ding{55}}%

\newcommand{\hl}[1]{\colorbox{yellow!20}{#1}}

\definecolor{lightgray}{gray}{0.9}
\newcommand{\llbox}[1]{%
	\vspace{4pt}%
	\noindent%
	\setlength{\fboxsep}{6pt}%
	\fcolorbox{black}{lightgray}{%
		\parbox{0.93\linewidth}{#1}%
	}%
}

%% file: cmd/definitions.tex
\def\arraystretch{1.2}
\newif\ifshowcomments

\makeatletter
\def\endthebibliography{%
	\def\@noitemerr{\@latex@warning{Empty `thebibliography' environment}}%
	\endlist
}
\makeatother

%% file: sections/abstract.tex
On average, 71\% of the code in typical Java projects comes from open-source software (OSS) dependencies, making OSS dependencies the dominant component of modern software code bases.
This high degree of OSS reliance comes with a considerable security risk of adding known security vulnerabilities to a code base.
To remedy this risk, researchers and companies have developed various \textit{dependency scanners}, which try to identify inclusions of known-to-be-vulnerable OSS dependencies.
However, there are still challenges that modern dependency scanners do not overcome, especially when it comes to dependency modifications, such as re-compilations, re-bundlings or re-packagings, which are common in the Java ecosystem.

To overcome these challenges, we present \jaralyzer, a \textit{bytecode-centric} dependency scanner for Java.
\jaralyzer does not rely on the metadata or the source code of the included OSS dependencies being available but directly analyzes a dependency's bytecode.

Our evaluation across 56 popular OSS components demonstrates that \jaralyzer outperforms other popular dependency scanners in detecting vulnerabilities within modified dependencies.
It is the only scanner capable of identifying vulnerabilities across all the above mentioned types of modifications.
But even when applied to unmodified dependencies, \jaralyzer outperforms the current state-of-the-art code-centric scanner \steady by detecting 28 more true vulnerabilities and yielding 29 fewer false warnings.

%% file: sections/introduction.tex
\section{Introduction}
\label{sec:introduction}

The use of open-source software (OSS) has become ubiquitous in modern software development.
OSS is now so prevalent in software projects that third-party dependencies account for the largest portion of the overall code base.
According to a 2023 report by Endor Labs on the state of dependency management~\cite{stateOfDependencyManagement2023}, an average of 71\% of the total code base in typical Java projects consists of OSS code.
This also means that the use of vulnerable OSS dependencies poses a real threat to modern software projects as additionally highlighted by the OWASP Top Ten ranking of the most critical application security risks~\cite{owaspTopTen}.
The infamous \textit{log4shell}~\cite{everson2022log4shell} and \textit{equifax breach}~\cite{luszcz2018apache} incidents, which had a major impact on the cybersecurity world, further emphasized the risk associated with vulnerable OSS.
To this end, various open-source~\cite{owaspDC, osvScanner, retireJS, ponta2020detection} but also commercial \textit{dependency scanners}~\cite{snykScanner, blackduckScanner, mendScanner, endorlabsScanner} have been developed, which seek to help minimize the posed threat by identifying known-to-be vulnerable OSS dependencies used within software projects.

\textit{Metadata-based} scanners conduct dependency scans by leveraging the metadata associated with dependency inclusions.
One common method involves generating a \textit{Software-Bill-of-Materials} (SBOM) for the project under analysis.
The SBOM lists all dependencies included in the software project along with their identifiers and versions.
This information is then compared against a security advisory of known vulnerable dependency versions, often represented as Common Vulnerability Enumeration (CVE) entries~\cite{cve}.
In the case of a Java Maven project this means creating an SBOM from the dependency information obtained within the project's \texttt{pom.xml} files and comparing it to some advisory.
However, as Ponta et al.~\cite{ponta2018beyond, ponta2020detection} have reported, this approach has multiple shortcomings.
For one, security advisories often do not contain precise information about which \textit{modules} of a specific OSS are affected by a certain CVE entry, causing metadata-based scanners to report false positives~\cite{ponta2020detection}.
E.g., if one considers the popular Spring framework for Java, advisories often assign a specific CVE entry to the whole framework, even though only one of its 20 modules is actually affected.
Even projects using only unaffected modules will nonetheless receive vulnerability alerts.
On the other hand, vulnerabilities and patches will not be reported by dependency scanners until the used advisories have been updated.
The median delay between public patch availability and advisory publication is 25 days~\cite{dependencyManagementReport2024}.
In the Maven ecosystem, the average delay extends to 41 days. 
This timeframe does not even account for the period between a fix becoming available in the project's source code repository and its eventual release to a public artifact repository.

To combat these issues Ponta et al.~\cite{ponta2020detection} have proposed a \emph{code-centric} approach, which does not rely on the information about known-to-be-vulnerable dependencies in advisories but directly compares the \emph{code} of included dependencies to a database of \emph{fix commits}.
These fix commits represent the real code changes performed within the OSS respective Git repositories that fix a specific CVE entry. 
Directly relying on the fix commits, one does not have to manually map the vulnerability to a list of affected OSS versions first.
Furthermore, having access to the exact code changes comprising the fix, one exactly knows which module of the OSS is affected by a vulnerability.
Moreover, it allows for a subsequent reachability analysis to check whether the vulnerable code segment is even executed within the context of the including software project.
This enables the approach to reduce false positives in the detection of vulnerabitilties when compared to metadata-based approaches.

Ponta et al. have implemented the code-centric approach in the dependency scanner \steady.
A prerequisite of the code-centric approach is that both fix commits and scanned dependencies are available in the same code representation to be comparable, an assumption that generally holds for languages like Python or JavaScript.
In contrast, Java poses a challenge, as dependencies are usually distributed in compiled \emph{bytecode}, whereas fix commits modify the original \emph{source code}.
To circumvent the issue of mismatched code representations, \steady attempts to retrieve the source code of dependencies from Maven Central, the most popular Java artifact repository. 
However, the source code may or may not be available alongside the compiled artifacts and, even when present, there is no guarantee that it actually matches the compiled artifact.
If \steady cannot retrieve the source code, it is unable to compare bytecode to source code~\cite{steadyDoc}, and thus requires costly and error-prone manual analysis to determine if a dependency is vulnerable.

While dependency scanners have gradually improved over the past years, important challenges remain, which considerably impair the detection performance of state-of-the-art tools.
Dann et al.~\cite{dann2021identifying} conducted a study evaluating the performance of six state-of-the-art dependency scanners---both open-source and commercial---on Java projects where dependencies were included in \textit{modified} forms.
These modifications include re-compilations, re-bundlings and re-packagings, which as determined by Dann et al. are widespread modifications in the Java ecosystem. 
In their investigation of a sample of 254 vulnerable classes, they identified 67,196 artifacts on Maven Central that contained these classes in a modified form, highlighting the prevalence of such modifications.
In these cases, the absence of precise metadata keeps metadata-based scanners from reliably detecting known-to-be-vulnerable dependencies, while \steady is hindered by the lack of source code resulting from most types of modifications.
A similar experiment conducted by Dietrich et al.~\cite{dietrich2023security} further confirmed these results and emphasized the prevalence of dependency modifications that can transparently be applied by the popular Maven Shade Plugin, unbeknown to developers.
Such modifications can even be found in the Java standard library.

To overcome these challenges, we propose \jaralyzer, our approach to \textit{bytecode-centric} Java dependency scanning, which extends the code-centric approach defined by Ponta et al.~\cite{ponta2020detection} to handle mismatched code representations.
\jaralyzer does not rely on the availability of metadata or source code, but directly compares a dependency's \emph{bytecode} against a fix-commit database.
Due to this capability, \jaralyzer does not require any manual analysis to determine if a dependency is vulnerable and is able to reliably detect known-to-be-vulnerable dependencies in modified forms.
To achieve this, \jaralyzer employs techniques to compile the code changes within a fix commit in isolation and normalizes the obtained bytecode to enable a comparison independent of the used compiler. 
Subsequently, it compares syntax, control and data flow to precisely identify whether known-to-be-vulnerable dependencies are included in the analyzed software project and whether corresponding fixes have been applied or not.

Our evaluation of \jaralyzer shows that, when it comes to detecting known-to-be-vulnerable dependencies included in modified form, it outperforms five other dependency scanners, four OSS and one commercial.
With a miss rate of at most 6\% for vulnerabilities in re-packaged dependencies compared to unmodified ones, it is the only scanner capable of handling all types of modifications identified by Dann et al~\cite{dann2021identifying}.
But even when applied to unmodified dependencies, when directly compared to the state-of-the-art code-centric dependency scanner \steady, \jaralyzer reported 28 more true and 29 fewer false vulnerabilities.

To summarize, the original contributions of the paper are:
\begin{itemize}
	\item A bytecode-centric Java dependency scanner, which does not rely on metadata or source code being available and is able to detect known-to-be-vulnerable OSS dependencies, even in modified form. 
	\item An implementation in an open-source tool\footnote{\url{https://github.com/stschott/jaralyzer}}.
	\item An evaluation across 56 popular Java OSS libraries using five different open-source and commercial dependency scanners.
\end{itemize}

%% file: sections/background.tex
\section{Background}
\label{sec:background}

In the following we introduce the fundamental topics of how Java dependencies are included and modified.

\subsection{Java Dependencies:}
\label{subsec:dependencyInclusion}
A Java project $P$ can be defined as a tuple $P = (C, D)$, where $C$ represents the application code written by the project’s developers, and $D$ denotes the set of \emph{dependencies}, i.e., third-party components originating from external sources. 
\jaralyzer focuses exclusively on analyzing the set of dependencies $D$, and does not examine the application code $C$.

The term \textit{dependency} refers to a separately distributed software library or framework that is included in a software project~\cite{dann2021identifying}.
In the case of Java, these software libraries are typically distributed as JAR archives containing the \textit{bytecode} of the library.
Bytecode is a machine-readable code representation Java source code is compiled to, allowing it to be executed by the Java Virtual Machine.
To manage the inclusion of dependencies, developers usually use build-automation tools such as Maven or Gradle.
In the case of Maven, the developer specifies the \texttt{groupId}, \texttt{artifactId} and \texttt{version} of the desired library within a \texttt{pom.xml} file, which contains the build information for the software project.
These three properties, referred to as \textit{GAV}, are used to uniquely identify a library.
During build time, Maven automatically retrieves the specified dependencies from configured artifact repositories, typically Maven Central.
In addition, Maven also resolves possible dependency version conflicts and retrieves transitive dependencies, which are the dependencies of the project's direct dependencies.

\subsection{Dependency Modifications:}
\label{subsec:dependecyModifications}
Dependencies in Java projects are not always distributed and included in the straightforward way described in Section~\ref{subsec:dependencyInclusion}.
Dann et al.~\cite{dann2021identifying} have identified four types of modifications developers frequently apply to OSS components, which are then in turn included by other developers as dependencies.
In the remainder of this paper, `modifications' refers to these four types.

\noindent
\textbf{Type 1 (patched):}
\label{subsubsec:type1}
This type of modification frequently occurs when developers fork the source code of an OSS and modify or patch it.
Dependencies modified this way do not include the original bytecode but the bytecode obtained when \textit{re-compiling} the modified source code.
Furthermore, one typically modifies the GAV by appending a suffix like \texttt{fix} or \texttt{patch} to it.

\noindent
\textbf{Type 2 (Uber-JAR):}
\label{subsubsec:type2}
In this type of modification, developers \textit{re-bundle} multiple OSS into a single JAR file, usually referred to as \textit{Uber-JAR}.
This type of modification is sometimes indicated by the JAR file containing \texttt{jar-with-dependencies} or \texttt{uber} in its file name.
In contrast to type 1, the original bytecode of each individual OSS is preserved.

\noindent
\textbf{Type 3 (bare Uber-JAR):}
\label{subsubsec:type3}
Type 3 modifications are similar to type 2 in the sense that multiple OSS are re-bundled into a single JAR file.
However, in contrast to type 2, the metadata contained within the JAR file (\texttt{pom.xml}, \texttt{META-INF} folder and file timestamps) is removed.
This type of modification can mostly be observed in legacy JAR files, which have been created before the advent of assembly plugins.

\noindent
\textbf{Type 4 (re-packaged Uber-JAR):}
\label{subsubsec:type4}
Type 4 modifications are also similar to type 2.
However, instead of just re-bundling multiple OSS into a single JAR file, the OSS are \textit{re-packaged}.
This involves modifying their original package names, either by prepending a string or by replacing them entirely with new names~\cite{mavenShadePluginRepackaging}.
This is often done to avoid name clashes and version conflicts.
In such cases the original bytecode of the OSS is changed substantially, because all references within the code need to be adjusted.

%% file: sections/concept.tex
\section{Bytecode-centric Dependency Scanning}
\label{sec:concept}

In the following we present \jaralyzer, a \textit{bytecode-centric} dependency scanning approach.

\input{sections/concept/overview.tex}
\input{sections/concept/knowledge-base-creation.tex}
\input{sections/concept/vulnerability-scanning.tex}
\input{sections/concept/re-packaging-detection.tex}

%% file: sections/concept/overview.tex
\subsection{Overview}
\label{subsec:overview}

\begin{figure}
	\includegraphics[width=\linewidth]{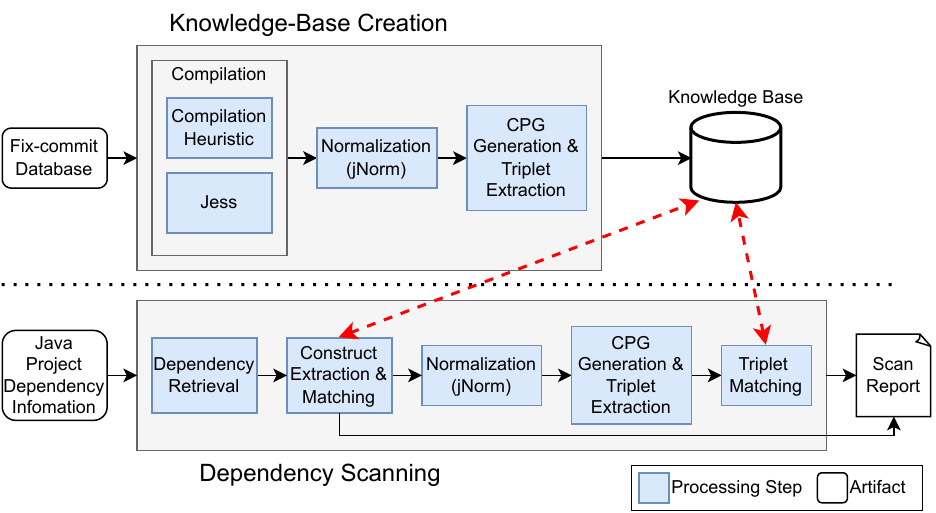}
	\caption{Overview of \jaralyzer}
	\label{fig:overview}
\end{figure}

Figure~\ref{fig:overview} shows an overview of \jaralyzer's architecture.
The approach consists of two major stages: \textit{(1)} the knowledge-base creation and the \textit{(2)} dependency scanning.
The constructed knowledge base serves as the foundation for the dependency scanning.
It contains all vulnerabilities that \jaralyzer can detect.

As a code-centric approach, \jaralyzer uses fix commits as input, which are commits fixing vulnerabilities in the source code repositories of OSS.
In the first step of the knowledge-base creation stage, \jaralyzer compiles the fix commits to bytecode using two techniques:
First it applies a compilation heuristic and, if this heuristic fails, it uses \jess~\cite{schott2024compilation} to compile commit changes within source code repositories.
After compilation of the fix commits, \jaralyzer normalizes the resulting bytecode using \textsc{jNorm}~\cite{schott2024java} to enable a subsequent comparison.
This step is necessary to remove differences that exist solely due to the use of different compilers. 
If not removed, those differences would interfere with matching the applied fix in the dependency scanning stage.
Afterwards, for each changed method in a given fix commit, \jaralyzer generates a code property graph (CPG)~\cite{yamaguchi2014modeling}, a representation that combines syntax, control flow, and data dependency information of a method.
Finally, \jaralyzer extracts comparable string triplets~\cite{bowman2020vgraph}, which encode edge information, from the generated CPGs and stores them in the knowledge base.
\jaralyzer uses these triplets to identify the presence of the fix in the next stage.

To scan a software project for known-to-be-vulnerable dependencies, \jaralyzer performs the steps described in the dependency scanning stage (see Figure~\ref{fig:overview}).
First, \jaralyzer uses the underlying build-automation tool of the project being scanned to extract all its dependencies.
\jaralyzer then scans each of the extracted JAR files individually for potential vulnerabilities.
To do so, \jaralyzer matches all \textit{constructs} (class, interface or method) within the JAR file to its knowledge base.
If \jaralyzer identifies a matching construct, it indicates that an included dependency in the software project may be vulnerable.
However, at this point, it remains unclear whether the version in use has already applied the fix.
To determine the presence of the fix, all constructs added, removed or changed in the fix are considered.
In particular, if the scanned construct is a method \textit{changed} in the fix---which is the most common scenario---\jaralyzer normalizes it, generates a CPG for it and transforms it into comparable string triplets.
Finally, \jaralyzer compares the triplets yielded from the scanned method's CPG to the triplets stored within its knowledge base.
Based on the degree of matched triplets, \jaralyzer reports the method as fixed or vulnerable.
This procedure is performed for each construct in each JAR file, resulting in a final report that contains all identified vulnerable dependencies.

In the following we will describe each processing step in detail.

%% file: sections/concept/knowledge-base-creation.tex
\subsection{Knowledge-Base Creation}
\label{subsec:knowledgeBaseCreation}

The knowledge-base creation stage is essential to code-centric dependency scanning.
In contrast to scanners relying on metadata, \jaralyzer requires the bytecode corresponding to vulnerability fixing commits.
This stage only needs to be executed once to create \jaralyzer's knowledge base, which is then available for every dependency scan.
After construction, for every CVE entry, the knowledge base contains the \textit{fully qualified name} (FQN) of each construct modified in the fix commits related to that CVE.
Each CVE entry contains at least one modified construct.
The FQN uniquely identifies a construct by specifying its full package and class name.
Additionally, the knowledge base contains each construct's modification type (added, removed, or changed), and the precise bytecode additions, removals and changes, extracted from the CPG as string triplets.

\subsubsection{Fix-commit Database}
\label{subsubsec:fixCommitDatabase}
As a bytecode-centric approach, \jaralyzer relies on a fix-commit database as input to create its knowledge base.
A fix-commit database maps CVE entries to the commits in an OSS project's source code repository that fix the respective CVE entries.
Possible fix-commit database options include SAP's \kb~\cite{ponta2019manually}, CVEFixes~\cite{bhandari2021cvefixes} or MoreFixes~\cite{akhoundali2024morefixes}.

\subsubsection{Compilation}
\label{subsubsec:compilation}
As dependencies are included in bytecode format but fix commits are in source code,  to enable a comparison \jaralyzer needs to generate bytecode for each construct modified by the fix commits of each CVE.
However, simply triggering a standard compilation with the configured build-automation tool is insufficient, as it frequently leads to failure~\cite{schott2024compilation, tufano2017there, hassan2017automatic, sulir2016quantitative, zhang2019large}.
As shown by Schott et al.~\cite{schott2024compilation}, performing a standard compilation following the approach of Tufano et al.~\cite{tufano2017there} successfully compiles only 14.9\% of CVE entries in the \kb fix-commit database.
However, for bytecode-centric dependency scanning one does not need to compile the entire code base, but only the source code files modified within the respective fix commit.
Leveraging this fact through a custom \emph{compilation heuristic} significantly improves the compilation success rate (see Section~\ref{subsec:setup}).
Instead of compiling the entire code base, \jaralyzer attempts to download all dependencies needed to compile the files modified in the fix commits and then compiles these files in isolation.
Second, if the heuristic fails, it applies the commit compilation tool \jess~\cite{schott2024compilation}.

\begin{figure}
	\includegraphics[width=\linewidth]{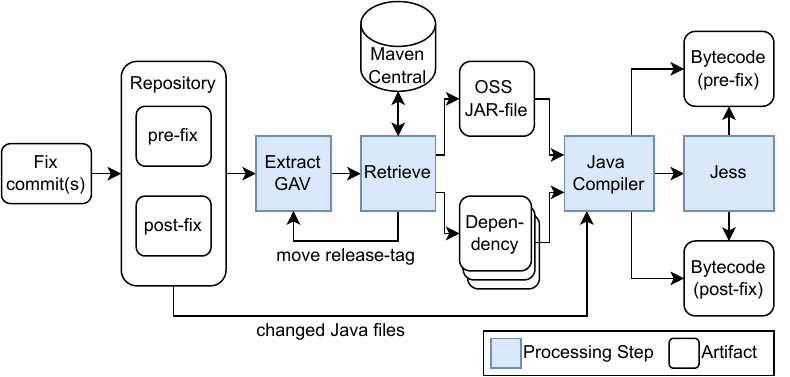}
	\caption{Detailed overview of \jaralyzer's compilation step}
	\label{fig:compilation}
\end{figure}

\begin{figure}
	\includegraphics[width=0.9\linewidth]{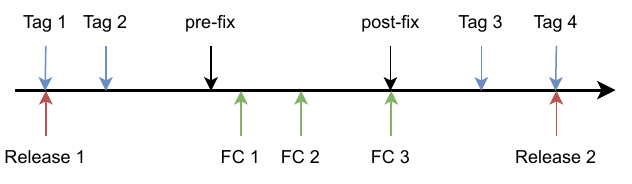}
	\caption{Exemplary version history of an OSS project with three fix commits. FC = Fix Commit}
	\label{fig:versionHistory}
\end{figure}

Figure~\ref{fig:compilation} shows a detailed overview of the compilation process.
It starts by cloning the source code repository corresponding to fix commits of a given CVE entry.
The fix might be contained within a single commit or segmented across multiple commits.
Figure~\ref{fig:versionHistory} shows an exemplary version history of a project with the fix consisting of the three fix commits FC1--FC3.
At first \jaralyzer collects all Java source code files (excluding test files) changed within the commits FC1--FC3.
To be able to extract the precise bytecode changes performed in the fix, \jaralyzer needs to compile the changed files before and after the fix is applied.
Thus, after collecting the changed files, \jaralyzer checks out the revision \textit{before} the first fix commit (FC1) is applied (see pre-fix in Figure~\ref{fig:versionHistory}).

To compile the changed files, we need to retrieve all necessary dependencies.
This includes a compiled release version of the OSS project, allowing us to avoid compiling the entire project and instead focus only on the changed files.
We do not use the compiled release version to extract the fix, as the original fix might have been further changed before the actual release and only use it as compilation supplement.
Directly using the release may interfere with the matching in modified cases, especially for type 1 modifications (see Section~\ref{subsubsec:type1}).
Here developers sometimes copy the original fix and manually apply it to their forked project.

To retrieve the compiled release version, \jaralyzer first tries to extract the GAV identifier from the project (see Figure~\ref{fig:compilation}).
This identifier is used to lookup if this specific version of the artifact exists on Maven Central.
If \jaralyzer does not find an artifact on Maven Central corresponding to the extracted coordinates, it will checkout the next repository revision tagged with a release tag and repeat the process.
Tags are typically used to mark release versions of the project, however sometimes developers tag a specific revision of the project that is not released on Maven Central.
Within the example shown in Figure~\ref{fig:versionHistory} one can see that Tag 3 is not pointing towards a release version of the project, but Tag 4 is.
Thus, \jaralyzer needs to move the tag at least two times to find a corresponding release on Maven Central.
We perform this step up to ten times, as our empirical testing revealed no cases where additional iterations resulted in a successful compilation.
If \jaralyzer finds a release version on Maven Central, it retrieves the corresponding JAR file and all available dependencies configured within the \texttt{pom.xml} file, which is hosted alongside the JAR file.
Extracting the GAV identifier typically works well for Maven projects, as one can simply invoke Maven to obtain this information.
For Gradle projects this is often not as trivial.
As Gradle build configurations are based on a Groovy/Kotlin-based language, Gradle first needs to execute the configuration scripts before the GAV can be extracted.
However, this execution often fails for outdated builds.
In such cases, we manually inspect the project's configuration scripts to extract the GAV and provide it to the compilation process.
Finally, using the retrieved dependencies, as well as the pre-compiled JAR file from Maven Central, \jaralyzer will forward the collected changed source code files from the repository, to a JDK Java compiler.
If the compilation is successful, \jaralyzer will obtain the bytecode of files changed in commits FC1--FC3 (see Figure~\ref{fig:versionHistory}), before the fix is applied.
Now to obtain the bytecode of the files \textit{after} the fix is applied, \jaralyzer will repeat the same process analogously for the post-fix version, by checking out the revision that commit FC3 is pointing to, as depicted in Figure~\ref{fig:versionHistory}.
If the compilation via the heuristic fails, \jaralyzer will invoke \jess~\cite{schott2024compilation}, a tool that has been specifically designed for the purpose of compiling commit changes within Java source code repositories in isolation.
Although, this is only a secondary option, as \jess might produce bytecode that is slightly different from the original bytecode and thus might cause interference for a subsequent comparison.
If \jaralyzer is able to compile only one of the pre- or post-fix revisions with the Java compiler, it will instead attempt to compile both revisions with \jess to ensure a consistent bytecode representation.

\subsubsection{Normalization}
\label{subsubsec:normalization}
Obtaining the bytecode of fix commits alone is not sufficient for detecting the presence of vulnerable bytecode in included dependencies.
Since we do not know what compilation environment the original dependencies have been compiled in, and different Java compilation environments (e.g. different compilers, versions and settings) produce different bytecode for equal source code~\cite{schott2024java, dann2019sootdiff}, the bytecode produced by \jaralyzer may differ from the one included in the scanned dependencies.
Since vulnerability fixes are often small, involving only a few lines of code, compilation differences may well mask the original fix and make a comparison impossible.
To make the bytecode comparable, we apply \textsc{jNorm}~\cite{schott2024java} right after compilation.
\textsc{jNorm} is a tool to normalize bytecode by removing nearly all differences introduced by different compilation environments.
After applying \textsc{jNorm}, compilation differences are mostly removed and the bytecode can be compared.

\subsubsection{CPG Generation \& Triplet Extraction}
\label{subsubsec:cpgGeneration}
After normalization, \jaralyzer uses the \textsc{SootUp}~\cite{karakaya2024sootup} static analysis framework to generate Code Property Graphs (CPG) for each method changed in the fix commits.
A CPG~\cite{yamaguchi2014modeling} is a graph-based representation of an individual method that combines syntax information in form of an abstract syntax tree, control flow information in form of a control flow graph and data, as well as control dependencies in form of a program dependence graph.
To extract the precise code instructions that comprise the fix, one must compute the differences between the CPG of the method before the fix was applied and the CPG of the method after the fix was applied.
This requires determining subgraph isomorphisms---a problem that is NP-complete---which makes a direct comparison of CPGs computationally expensive~\cite{bowman2020vgraph}.
We thus resort to an approximation proposed by Bowman and Huang~\cite{bowman2020vgraph} and extract graph \textit{triplets}.
For every edge $e$ within a given CPG $g$, a triplet $t$ is defined as $(n_s, e_l, n_t)$ where $e_l$ denotes the label of edge $e$, $n_s$ denotes the source node label and $n_t$ the target node label of edge $e$ within graph $g$.
The set $T$ is the set of all triplets for a given CPG $g$.
Once we have created a CPG $g_{vul}$ and a CPG $g_{fix}$ for a method before and after the fix has been applied we extract the respective triplet sets $T_{vul}$ and $T_{fix}$.
We then define the \textit{context triplets} $CT$, \textit{positive triplets} $PT$, and \textit{negative triplets} $NT$:
\begin{align*}
CT = T_{vul} \cap T_{fix} \qquad
PT = T_{fix} \setminus T_{vul} \qquad
NT = T_{vul} \setminus T_{fix}
\end{align*}
$CT$ contains all triplets that have not been changed by the fix, $PT$ contains triplets that have been added by the fix and $NT$ contains triplets that have been removed by the fix.
Thus, $PT$ and $NT$ contain the fine-grained code changes performed within the fix.
Finally, we store the generated triplet sets for each method in \jaralyzer's knowledge base, alongside the information about the FQN of the method and the affecting CVE entry.

%% file: sections/concept/vulnerability-scanning.tex
\subsection{Dependency Scanning}
\label{subsec:vulnerabilityScanning}

In the dependency scanning stage, \jaralyzer is applied to a Java software project containing dependency information and tries to identify whether known-to-be-vulnerable dependencies are included.

\subsubsection{Dependency Retrieval}
\label{subsubsec:dependencyExtraction}

\jaralyzer starts with retrieving all dependencies included in the scanned project.
Depending on the used build-automation tool, it invokes a command of the tool (e.g. the \texttt{copy-dependencies} goal of the maven-dependency-plugin~\cite{mavenDependencyPlugin}) to retrieve the JAR files corresponding to each dependency as defined in the project's dependency information (e.g. \texttt{pom.xml} for Maven projects).

\subsubsection{Construct Extraction \& Matching}
\label{subsubsec:constructExtraction}

In this step, \jaralyzer individually processes each previously retrieved dependency JAR file, first extracting the FQN of every construct within it.
Then, \jaralyzer queries the knowledge base with each extracted FQN to verify if any of these FQNs is associated with a vulnerability.
This query results in a set of CVE entries potentially affecting the JAR file.
\jaralyzer proceeds to scan the JAR file for each potential CVE entry individually.
Here it distinguishes between constructs removed, added or changed by the fix:
\textit{(1)} If, within the scanned JAR file, a construct is detected that is marked as \textit{removed} by the fix of the respective CVE entry, \jaralyzer classifies the corresponding construct as vulnerable, otherwise as fixed.
\textit{(2)} If a method is marked as \textit{added} by the fix but is missing from the scanned JAR file while its declaring class is present, \jaralyzer classifies the construct as vulnerable, otherwise as fixed.
\textit{(3)} If \jaralyzer detects a method in the scanned JAR file that has been \textit{changed} by the fix, it extracts the method body and proceeds with normalization (see Section~\ref{subsubsec:normalization}) and CPG generation (see Section~\ref{subsubsec:cpgGeneration}), following the same process used in the knowledge-base creation stage.

\subsubsection{Triplet Matching}
\label{subsubsec:tripletMatching}
After generating a CPG $g_{m}$ and extracting the set of triplets $T_{m}$ for the matched and normalized method (see Section~\ref{subsubsec:cpgGeneration}), \jaralyzer performs a triplet matching to determine whether the method is in a vulnerable or fixed state.
To do so, it compares the triplet set $T_{m}$ to the set of positive triplets $PT$ and negative triplets $NT$ of the matched method within the knowledge base.
If the equation $|NT \cap T_{m}| \geq |PT \cap T_{m}|$ holds, it indicates that the matched method is more or equally similar to the method \textit{before} the fix has been applied. 
\jaralyzer thus classifies the matched method as vulnerable.
If the fix did not remove or change any code, but just added code within a method, the set of negative triplets $NT$ is always empty.
In such cases, \jaralyzer verifies, whether $|PT \cap T_{m}| \ge \theta_{PT}$, for a configurable threshold $\theta_{PT}$.

In the end, \jaralyzer counts these matched constructs that are associated with a specific vulnerability and have been classified as vulnerable.
If the number of constructs classified as vulnerable is greater than or equal to those classified as fixed, \jaralyzer reports the included JAR file as vulnerable to this specific CVE entry.
Especially when the initial fix is changed over time, it might be the case that some constructs are reverted to their pre-fix state and \jaralyzer thus classifies some constructs as fixed and others as vulnerable.
In cases involving discrepancies, tools such as \steady require human intervention to reach a decision. 
In contrast, because \jaralyzer is designed to minimize human involvement, we decided on this majority criterion, which yielded the best performance in empirical testing.
Finally, \jaralyzer continues to process the next CVE entry potentially affecting the JAR file.

After scanning all provided JAR files, \jaralyzer generates a final scan report that lists all discovered vulnerable dependencies along with their associated CVE entries.

%% file: sections/concept/re-packaging-detection.tex
\subsection{Re-Packaging Detection}
\label{subsec:repackagingDetection}

\jaralyzer has been designed to not require any metadata about included dependencies within its dependency scanning stage.
This means that, by construction, it does not impact \jaralyzer's detection capabilities whether the included dependencies are re-compiled, re-bundled or their associated metadata is removed (type 1--3 modifications).
As long as the bytecode is available, \jaralyzer will be able to scan them for known-to-be-vulnerable OSS.
However, to uniquely identify constructs and to initially match them to its knowledge base, \jaralyzer does rely on their FQNs.
The FQNs of constructs, however, change when the dependency is re-packaged (type 4 modification, see Section~\ref{subsubsec:type4}).
Listing~\ref{lst:repackagingExample} shows a source code example (for illustration purposes) of such a re-packaging, which is typically applied to bytecode using tools like the Maven Shade Plugin~\cite{dietrich2023security}.
Listing~\ref{lst:original} shows the original class, while Listing~\ref{lst:repackaged} shows the same class, but re-packaged.
The comments in lines~4 and 7, above class \texttt{C} and method \texttt{foo}, show their respective FQNs.
Differences between both listings are highlighted.
Due to re-packaging, which prepends the string ``r'' in front of each package name within the dependency, the FQNs of all constructs change and will never match with the knowledge base in case a vulnerability is associated with method \texttt{foo} in Listing~\ref{lst:original}.
To cover type 4 modifications, \jaralyzer uses a \textit{re-packaging detection} mode that does not rely on FQNs for matching.
The re-packaging detection follows the same process shown in Figure~\ref{fig:overview}, yet instead of relying on FQNs during the construct-matching step, it uses the \textit{unqualified} names of constructs (class names \textit{without} package names).
E.g., instead of querying the knowledge base for the fully qualified method signature \texttt{r.a.C: void foo(r.a.b.X)} (see Listing~\ref{lst:repackaged}), \jaralyzer will query for all methods in the knowledge base having the \textit{unqualified} signature \texttt{C: void foo(X)}.
As the unqualified signature does no longer uniquely identify methods, \jaralyzer additionally checks for the \textit{class context} to avoid spurious matches: 
it checks whether all sibling methods and fields of the scanned method are also present in the method matched in the knowledge base.
The information about the class context has also been collected during the knowledge-base creation stage.
As an example, if the method name \texttt{foo} from Listing~\ref{lst:repackaged}) matches the knowledge base, the class context will consider the sibling field \texttt{baz} (line~6 in Listing~\ref{lst:repackaged}).
If the class context exceeds a configurable threshold $\theta_{CC}$, \jaralyzer considers the method as a correct match.

Finally, the triplet matching step also slightly differs from \jaralyzer's default process.
As in Java bytecode type names are always fully resolved, \jaralyzer has to unqualify all triplets.
E.g., consider the call of method \texttt{bar} in line~9 of Listing~\ref{lst:repackaged}.
Within the bytecode, and thus within its corresponding triplet, the method call will be expressed with the fully qualified name of \texttt{X} (i.e., \texttt{r.a.b.X}).
Thus, before matching, we unqualify all triplets within the scanned JAR and the knowledge base.
To avoid false-positive matches, before comparing the triplet set of the scanned method $T_m$ to the set of positive and negative triplets, we verify that the equation $|CT \cap T_m| > \theta_{CT}$ holds for a configurable threshold $\theta_{CT}$.
Doing this, we verify whether the part of the method that has \textit{not} been changed by the fix, expressed by the set of context triplets $CT$, is sufficiently similar to the scanned method.

These adjustments however, come with the cost of potentially missing vulnerable dependencies or erroneously reporting a dependency as vulnerable, often depending on the choice of the configurable thresholds.
Thus, depending on the use case, the default mode and the re-packaging detection mode can be run in tandem or independent of each other.
When only running the default mode, one will not identify any re-packaged known-to-be-vulnerable dependencies, while only running the re-packaging detection mode, one might receive more false-positive vulnerability reports.

Although re-packaging detection could be added as a step within \jaralyzer's main pipeline, we added it as a parallel mode to avoid repeated, costly executions of the full pipeline, including normalization and CPG-generation, caused by ambiguous method signatures.
By providing two independent modes, developers can either run both or choose the one that best suits their specific needs.

\setcounter{figure}{0}
\renewcommand{\figurename}{Listing}
\noindent
\begin{figure}
	\begin{minipage}[]{.45\linewidth}
		\begin{lstlisting}
package a;
import a.b.X;

// a.C
class C {
  int baz;
  // a.C: void foo(a.b.X)
  void foo(X x) {
    int i = x.bar();
  } 
}
		\end{lstlisting}
		\subcaption{Original class}
		\label{lst:original}
	\end{minipage}
	\hfill
	\begin{minipage}[]{.51\linewidth}
		\begin{lstlisting}
package §\hl{r.}§a;
import §\hl{r.}§a.b.X;

// §\hl{r.}§a.C
class C {
  int baz;
  // §\hl{r.}§a.C: void foo(§\hl{r.}§a.b.X)
  void foo(X x) {
    int i = x.bar();
  } 
}
		\end{lstlisting}
		\subcaption{Re-packaged class}
		\label{lst:repackaged}
	\end{minipage}
	\caption{Re-packaging example}
	\label{lst:repackagingExample}
\end{figure}
\renewcommand{\figurename}{Figure}
\setcounter{figure}{3}

%% file: sections/evaluation.tex
\newcommand{\RQONE}{How does \jaralyzer compare to state-of-the-art dependency scanners in identifying \textit{modified} known-\allowbreak to-\allowbreak be-\allowbreak vulnerable dependencies?}
\newcommand{\RQTWO}{How does \jaralyzer compare to the state-of-the-art code-centric dependency scanner in identifying \textit{unmodified} known-\allowbreak to-\allowbreak be-\allowbreak vulnerable dependencies?}
\newcommand{\RQTHREE}{How runtime-efficient is \jaralyzer?}

\section{Evaluation}
\label{sec:evaluation}

In the following we evaluate the effectiveness of \jaralyzer.
To do so we answer the following research questions.

\begin{enumerate}[rightmargin=0.9em]
	\item[\textbf{RQ1:}] \RQONE
	\item[\textbf{RQ2:}] \RQTWO
	\item[\textbf{RQ3:}] \RQTHREE
\end{enumerate}

The first research question focuses on the challenge that current dependency scanners face, when dependencies are included in modified form.
The second research question evaluates \jaralyzer's detection performance in the general case, when dependencies are not modified.
To do so, we compare \jaralyzer against the state-of-the-art code-centric dependency scanner \steady.
The third research question evaluates \jaralyzer's runtime efficiency.

\input{sections/evaluation/setup.tex}
\input{sections/evaluation/rq1.tex}
\input{sections/evaluation/rq2.tex}
\input{sections/evaluation/rq3.tex}

%% file: sections/evaluation/setup.tex
\subsection{Experimental Setup}
\label{subsec:setup}

\begin{figure}
	\includegraphics[width=\linewidth]{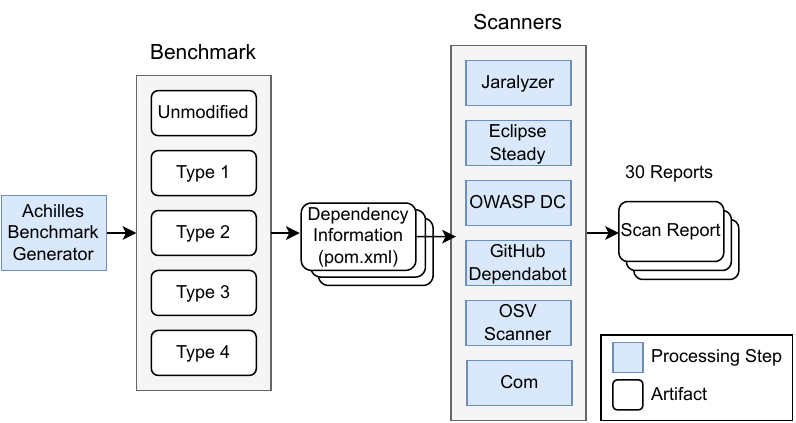}
	\caption{Overview of the experimental setup}
	\label{fig:experimentalSetup}
\end{figure}

Figure~\ref{fig:experimentalSetup} shows an overview of our experimental setup.
The evaluation is based on the \achilles benchmark generator created by Dann et al.~\cite{dann2021identifying}.
\achilles comes with a dataset of 534 distinct OSS artifacts and allows for an automatic creation of Maven projects having dependencies on a selected subset of these artifacts, either in unmodified or modified form (type 1--4).
Most of the provided artifacts are different versions of the same OSS.
We only included the latest provided version\footnote{For OSS comprised of multiple modules, e.g. Spring and Jetty, we did not include the latest version provided in \achilles but the version with the most compatible modules}
of each distinct OSS (unique groupId and artifactId), as including multiple versions of the same OSS within a single project would lead to version conflicts.
Using the \achilles benchmark generator we created five distinct Maven projects, each having 56 artifacts as dependencies, including popular OSS such as Guava, Spring and Jackson-Databind.
According to Endor Labs~\cite{dependencyManagementReport2024}, these are among the Java OSS accounting for most vulnerability findings.
We kept the 56 artifacts together in a single project, as splitting them should not impact tool behavior and would require introducing arbitrary splitting criteria.
Based on the definition presented in Section~\ref{subsec:dependencyInclusion}, we create five distinct projects $P = (C, D)$, with $C$ being empty as no application code is provided and $D$ containing the 56 artifacts from \achilles as dependencies.
While one of the created projects contained the dependencies in unmodified form, the other four projects contained \textit{modified} dependencies according to types 1--4 (see Section~\ref{subsec:dependecyModifications}).
We then provided the dependency information (\texttt{pom.xml} files) belonging to each of the generated projects to six different dependency scanners: \jaralyzer, \steady~\cite{ponta2020detection}, \owaspDC~\cite{owaspDC}, GitHub Dependabot~\cite{dependabot}, OSV Scanner~\cite{osvScanner} and a commercial dependency scanner (Com).
Applying each tool to the five different projects within our benchmark, we obtained a total of 30 scan reports containing the CVE entries that each tool detected for the individual projects.

As \jaralyzer's knowledge base, we use SAP's \kb~\cite{ponta2019manually}, as it is also the knowledge base utilized by \steady.
It contains a total of 1,297 manually-curated CVEs from 2005--2023, affecting popular and industry-relevant Java and Python OSS.
We filtered out all entries that do not affect Java OSS or include no changes to Java source code files from \kb. 
This left us with 733 relevant entries.
\jaralyzer was able to compile 542 of these 733 entries using its compilation heuristic (see Section~\ref{subsubsec:compilation}).
For 170 of these compilations we had to supplement additional information about GAVs.
\textsc{Jess}~\cite{schott2024compilation} was able to compile 122 of the remaining entries (54 full and 68 partial compilations).
Thus, leveraging the fact that only files modified in the fix commits need to be compiled, \jaralyzer was able to compile 664/733 (90.6\%) entries of \kb for Java OSS. 
We also performed an investigation of the failed cases.
Fix commits are applied in-between two releases (see Figure~\ref{fig:versionHistory}).
Whenever there are code dependencies towards the release versions before \textit{and} after the commits, the compilation fails, as we can only provide one of the versions for compilation.
Furthermore, in few cases the compiled release version or the dependencies of the project to be compiled are not stored on Maven Central but on different repositories.

In total, leveraging the fact that only few files need to be compiled, \jaralyzer achieves a comparatively high compilation success rate on Java repository snapshots~\cite{hassan2017automatic, sulir2016quantitative, zhang2019large}, bringing the 14.9\% success rate reported by Schott et al.~\cite{schott2024compilation} when applying the procedure of Tufano et al.~\cite{tufano2017there} on \kb, up to 90\%. 
Nevertheless, there remains room for improvement to enable the detection of all CVE entries within the \kb fix-commit database.

All dependency scanners were executed within a Docker container (v26.1.4) running Alpine Linux v3.19 with Eclipse Temurin JDK 17.
The system running the container, was configured to use four cores of an Intel Xeon E5-2695 v3 (2.3 GHz) CPU and 32GB of RAM.
Furthermore, we used \owaspDC v11.1.0, OSV Scanner v1.4.3 and \steady v3.2.5.
GitHub Dependabot was executed in December 2024. 
We did not execute \steady's reachability analysis, as our evaluation focuses on comparing detection performance.
\steady classifies certain reported vulnerabilities as ``unknown'', requiring manual analysis (57 in the unmodified benchmark, all of them in the modified benchmarks).
In fact, for 4 out of the 56 artifacts in the unmodified benchmark, the source code is not available on Maven Central and for the modified benchmarks, no source code is available.
We count these cases as reported vulnerabilities. 
It is worth noting that, even when source code is available, there is no guarantee that it is accurate: Maven Central does not verify if the source code matches the actual bytecode and even allows uploading placeholder files that contain no source code at all~\cite{mavenCentralSources}.
Another interesting observation we made is that \steady performs significantly worse when \achilles packages the artifacts on Windows compared to Linux, so we let \achilles package all artifacts on a Linux machine.
Finally, we executed \jaralyzer in default and re-packaging detection mode and configured the (default) threshold values: $\theta_{PT}$ = 0.5, $\theta_{CC}$ = 0.3, $\theta_{CT}$ = 0.3.
Through empirical testing, we found these values to provide the best trade off between detecting re-packaged known-to-be-vulnerable dependencies and minimizing wrong alerts.

%% file: sections/evaluation/rq1.tex
\subsection{RQ1: \RQONE}
\label{subsec:rqone}

In this research question we evaluate \jaralyzer's performance on \textit{modified} dependencies, which pose a major challenge for current dependency scanners~\cite{dann2021identifying, dietrich2023security}.
To measure the effectiveness of dependency scanners on modified dependencies, Dann et al. used a small benchmark of seven distinct artifacts created with the \achilles benchmark generator, applying the same modifications as we did in our experimental setup (see Section~\ref{subsec:setup}).
To determine whether each tool reported a true or false positive, they rely on a ground truth, which they have semi-automatically uncovered within their~\cite{dann2021identifying} study and added to the \achilles benchmark generator.
We manually inspected the ground truth used in their experiment and uncovered that it is not fully accurate.
Even in their small benchmark containing only 7 distinct artifacts and 15 distinct CVE entries, two of the test cases are incorrect.
They report httpclient 4.1.3 as being affected by CVE-2015-5262, whereas it only affects httpclient as of version 4.3\footnote{\url{https://issues.apache.org/jira/browse/HTTPCLIENT-1478}}.
Furthermore, they erroneously do not consider CVE-2018-11040 as affecting spring-webmvc 5.0.0.RELEASE\footnote{\url{https://security.snyk.io/vuln/SNYK-JAVA-ORGSPRINGFRAMEWORK-467268}}.
Because of these inaccuracies, we decided not to rely on the ground truth provided by \achilles.
Instead, we compare \jaralyzer head-to-head with each of the five other tools within our setup.
We assume that if a tool is unaffected by a specific type of modification, it should report the same CVE entries for both the unmodified and modified benchmarks.
First, we filter all the CVE entries where both tools \textit{agree} upon the respective CVE entry affecting the unmodified benchmark.
These agreed upon CVE entries are the only ones we consider for the modified benchmarks.
Then, we investigate how many of the same CVE entries reported for the unmodified benchmark by both tools are also reported for the type 1--4 benchmarks.

\begin{table}[]
	\centering
	\caption{Detected CVE entries by each scanner in comparison to \jaralyzer on modified dependency inclusions. Bold numbers indicate CVE entries missed by \jaralyzer.}
	\resizebox{0.95\linewidth}{!}{
		\begin{tabular}{l|ccccc}
			& \textbf{Agreed}     & \textbf{Type 1} & \textbf{Type 2} & \textbf{Type 3} & \textbf{Type 4} \\ \hline
			\textbf{Jaralyzer}   & \multirow{2}{*}{61} & 61              & \textbf{58}     & \textbf{58}     & \textbf{54}     \\
			\textbf{Steady}      &                     & 59              & 61              & 61              & -               \\ \hline
			\textbf{Jaralyzer}   & \multirow{2}{*}{78} & 78              & 78              & 78              & \textbf{74}     \\
			\textbf{OWASP DC}    &                     & 74              & 56              & -               & 56              \\ \hline
			\textbf{Jaralyzer}   & \multirow{2}{*}{57} & 57              & 57              & 57              & \textbf{56}     \\
			\textbf{Dependabot}  &                     & 56              & -               & -               & -               \\ \hline
			\textbf{Jaralyzer}   & \multirow{2}{*}{64} & 64              & 64              & 64              & \textbf{62}     \\
			\textbf{OSV Scanner} &                     & -               & -               & -               & -               \\ \hline
			\textbf{Jaralyzer}   & \multirow{2}{*}{72} & 72              & 72              & 72              & \textbf{71}     \\
			\textbf{Com}         &                     & 72              & -               & -               & -              
		\end{tabular}
	}
	\label{tab:modificationResults}
\end{table}

Table~\ref{tab:modificationResults} presents the results of this experiment.
It shows the head-to-head comparison of \jaralyzer with each of the tools and the number of agreed upon CVE entries for the unmodified benchmark (column ``CVEs'').
Columns ``Type 1'' to ``Type 4'' show how many of agreed upon CVE entries are reported by each tool for the type 1--4 benchmarks.
None of the tools \jaralyzer is compared against is able to handle all types of modifications.
While some tools perform better on certain types of modifications, e.g., \steady performing well on types 1--3, GitHub Dependabot, OSV Scanner and the commercial tool completely fail for type 2--4 modifications.
Throughout type 1--3 modifications, \jaralyzer does not miss any CVE entry, except for three in its comparison to \steady.
A detailed investigation of these entries reveals this to be an interesting case, where the three missed CVE entries are actually \textit{false positives} for the unmodified benchmark reported by both \jaralyzer and \steady.
These false positives are caused by the code-centric nature of the tools.
The corresponding CVE entries (CVE-2013-6429, CVE-2014-0225, CVE-2015-2080) are affecting the multi-module projects Spring and Jetty.
If the fix commits associated with a CVE entry modify code across multiple modules of a project, even though the vulnerability only affects a single module, this may interfere with detection in certain cases.
For example, CVE-2015-2080 affects the \textit{jetty-http} module of the Jetty project, however, its fix commits modify code in both the \textit{jetty-http} and \textit{jetty-util} modules.
Now, if the JAR files corresponding to jetty-http and jetty-util are scanned individually, both tools erroneously report jetty-util as being vulnerable, since the majority of the fix code, which is applied to jetty-http, cannot be identified within jetty-util.
However, in type 2--4 modifications, all dependencies are re-bundled into a single JAR-file.
This then allows the tools to match all code changes from the respective fix commits at once.
In contrast to \steady, \jaralyzer is able to take advantage of this, and correctly determines that the re-bundled JAR file is unaffected by these CVE entries, reducing the signaled entries from 61 to 58 (54 for type 4).
An important point to consider is that, due to the unavailability of source code, all CVE entries flagged by \steady in the modified benchmarks, including those not agreed to by \jaralyzer, are classified as ``unknown'', meaning that \steady requires manual analysis for each case.

None of the compared tools is able to handle type 4 modifications, except for \owaspDC, which still misses 22 CVE entries.
For type 4 modifications, \jaralyzer, even though it does not find all CVE entries it detected in the other benchmarks, considerably outperforms the other tools.
The missed CVE entries are caused by not being able to rely on the FQNs and thus \jaralyzer having to rely on its re-packaging detection mode with the configured thresholds.
To further assess whether modifications affect \jaralyzer's \textit{precision}, we examined whether it reported CVE entries for type 1–-4 modifications that were not reported for the unmodified benchmark.
This did not occur.

While our results show that no tool within our experimental setup, except \jaralyzer, is able to effectively handle modified dependencies, we are not fully able to confirm the results of Dann et al.
Regarding the ability to handle different types of modifications, our results align with those of Dann et al., with e.g. \owaspDC being the only tool to partially support type 4 modifications.
However, unlike their findings, we do not observe the same significant performance degradation for modifications that the tools can still partially handle (e.g., types 1–3 for \steady), even when accounting for the inaccurate ground truth.

Finally, we evaluated \jaralyzer on the small benchmark (with adjusted ground truth) used within Dann et al.'s study.
\jaralyzer achieved 100\% recall and precision across all modification types.

\llbox{
	Among the evaluated tools, \jaralyzer is the only one capable of handling all types of modified dependency inclusions, missing at most 6\% of vulnerabilities in type 4 modifications.
}

%% file: sections/evaluation/rq2.tex
\subsection{RQ2: \emph{Unmodified} Dependencies Evaluation}
\label{subsec:rqtwo}

This research question considers the scenario where \textit{no} modifications are applied to the dependency inclusions.
To evaluate \jaralyzer's detection performance, we directly compare it to the state-of-the-art code-centric dependency scanner \steady.
We perform a study similar to the one conduced by Ponta et al.~\cite{ponta2020detection}, where they compare the findings of \steady to the findings of \owaspDC and manually review the cases where the tools disagree.
In their conducted study they determine that \steady significantly outperforms \owaspDC and detects considerably fewer false positives while also finding more true positives.
Thus, we compare \jaralyzer to \steady.

As reported in Section~\ref{subsec:rqone}, there are 61 CVE entries that both tools report and agree upon for the unmodified benchmark.
In this RQ we primarily focus on the cases where the tools \textit{disagree}.
We only considered the 664 CVE entries shared across the respective knowledge bases of \jaralyzer and \steady.

\begingroup
\setlength\tabcolsep{2 pt}
\renewcommand{\arraystretch}{0.9}
\begin{table}[]
	\caption{Disjunct CVE reports of \jaralyzer and \steady for the unmodified benchmark \\ \centerline{\textcolor{Green}{\cmark} = true positive; \textcolor{Red}{\xmark} = false positive}}
	\begin{subtable}{.456\linewidth}
		\caption{\jaralyzer reports}
		\resizebox{\linewidth}{!}{
			\begin{tabular}{lcccc}
				\multicolumn{1}{l|}{\textbf{CVE}}   & \multicolumn{1}{c|}{\textbf{KB}} & \multicolumn{1}{c|}{\textbf{GA}} & \multicolumn{1}{c|}{\textbf{NVD}} & \textbf{SV} \\ \hline
				\rowcolor{gray!25}
				\multicolumn{5}{l}{commons-fileupload-1.3.2}                                                                                                                \\
				\multicolumn{1}{l|}{CVE-2016-6793\footnote{CVE actually affects Apache Wicket, however, affected file is copied from commons-fileupload}}  & \multicolumn{1}{c|}{-}           & \multicolumn{1}{c|}{\color{Red}{\xmark}}          & \multicolumn{1}{c|}{\color{Red}{\xmark}}           & \color{Red}{\xmark}          \\
				\rowcolor{gray!25}
				\multicolumn{5}{l}{dom4j-1.6.1}                                                                                                                             \\
				\multicolumn{1}{l|}{CVE-2020-10683} & \multicolumn{1}{c|}{-}           & \multicolumn{1}{c|}{\color{Green}{\cmark}}          & \multicolumn{1}{c|}{\color{Green}{\cmark}}           & \color{Green}{\cmark}            \\
				\rowcolor{gray!25}
				\multicolumn{5}{l}{itextpdf-5.5.0}                                                                                                                          \\
				\multicolumn{1}{l|}{CVE-2017-9096}  & \multicolumn{1}{c|}{-}           & \multicolumn{1}{c|}{\color{Green}{\cmark}}          & \multicolumn{1}{c|}{\color{Green}{\cmark}}           & \color{Green}{\cmark}          \\ 
				\rowcolor{gray!25}
				\multicolumn{5}{l}{jackson-databind-2.9.7}                                                                                                                  \\
				\multicolumn{1}{l|}{CVE-2019-12086} & \multicolumn{1}{c|}{\color{Green}{\cmark}}          & \multicolumn{1}{c|}{\color{Green}{\cmark}}          & \multicolumn{1}{c|}{\color{Green}{\cmark}}           & \color{Green}{\cmark}           \\
				\multicolumn{1}{l|}{CVE-2019-12384} & \multicolumn{1}{c|}{\color{Green}{\cmark}}          & \multicolumn{1}{c|}{\color{Green}{\cmark}}          & \multicolumn{1}{c|}{\color{Green}{\cmark}}           & \color{Green}{\cmark}           \\
				\multicolumn{1}{l|}{CVE-2019-12814} & \multicolumn{1}{c|}{\color{Green}{\cmark}}          & \multicolumn{1}{c|}{\color{Green}{\cmark}}          & \multicolumn{1}{c|}{\color{Green}{\cmark}}           & \color{Green}{\cmark}           \\
				\multicolumn{1}{l|}{CVE-2019-14379} & \multicolumn{1}{c|}{\color{Green}{\cmark}}          & \multicolumn{1}{c|}{\color{Green}{\cmark}}          & \multicolumn{1}{c|}{\color{Green}{\cmark}}           & \color{Green}{\cmark}           \\
				\multicolumn{1}{l|}{CVE-2019-14439} & \multicolumn{1}{c|}{\color{Green}{\cmark}}          & \multicolumn{1}{c|}{\color{Green}{\cmark}}          & \multicolumn{1}{c|}{\color{Green}{\cmark}}           & \color{Green}{\cmark}          \\
				\multicolumn{1}{l|}{CVE-2019-14892} & \multicolumn{1}{c|}{\color{Green}{\cmark}}          & \multicolumn{1}{c|}{\color{Green}{\cmark}}          & \multicolumn{1}{c|}{\color{Green}{\cmark}}           & \color{Green}{\cmark}           \\
				\multicolumn{1}{l|}{CVE-2019-14893} & \multicolumn{1}{c|}{\color{Green}{\cmark}}          & \multicolumn{1}{c|}{\color{Green}{\cmark}}          & \multicolumn{1}{c|}{\color{Green}{\cmark}}           & \color{Green}{\cmark}           \\
				\multicolumn{1}{l|}{CVE-2019-16942} & \multicolumn{1}{c|}{\color{Green}{\cmark}}          & \multicolumn{1}{c|}{\color{Green}{\cmark}}          & \multicolumn{1}{c|}{\color{Green}{\cmark}}           & \color{Green}{\cmark}           \\
				\multicolumn{1}{l|}{CVE-2019-16943} & \multicolumn{1}{c|}{\color{Green}{\cmark}}          & \multicolumn{1}{c|}{\color{Green}{\cmark}}          & \multicolumn{1}{c|}{\color{Green}{\cmark}}           & \color{Green}{\cmark}           \\
				\multicolumn{1}{l|}{CVE-2019-17267} & \multicolumn{1}{c|}{\color{Green}{\cmark}}          & \multicolumn{1}{c|}{\color{Green}{\cmark}}          & \multicolumn{1}{c|}{\color{Green}{\cmark}}           & \color{Green}{\cmark}           \\
				\multicolumn{1}{l|}{CVE-2019-17531} & \multicolumn{1}{c|}{\color{Green}{\cmark}}          & \multicolumn{1}{c|}{\color{Green}{\cmark}}          & \multicolumn{1}{c|}{\color{Green}{\cmark}}           & \color{Green}{\cmark}           \\
				\multicolumn{1}{l|}{CVE-2019-20330} & \multicolumn{1}{c|}{\color{Green}{\cmark}}          & \multicolumn{1}{c|}{\color{Green}{\cmark}}          & \multicolumn{1}{c|}{\color{Green}{\cmark}}           & \color{Green}{\cmark}           \\
				\multicolumn{1}{l|}{CVE-2020-8840}  & \multicolumn{1}{c|}{\color{Green}{\cmark}}          & \multicolumn{1}{c|}{\color{Green}{\cmark}}          & \multicolumn{1}{c|}{\color{Green}{\cmark}}           & \color{Green}{\cmark}           \\
				\multicolumn{1}{l|}{CVE-2020-9546}  & \multicolumn{1}{c|}{\color{Green}{\cmark}}          & \multicolumn{1}{c|}{\color{Green}{\cmark}}          & \multicolumn{1}{c|}{\color{Green}{\cmark}}           & \color{Green}{\cmark}           \\
				\multicolumn{1}{l|}{CVE-2020-9547}  & \multicolumn{1}{c|}{\color{Green}{\cmark}}          & \multicolumn{1}{c|}{\color{Green}{\cmark}}          & \multicolumn{1}{c|}{\color{Green}{\cmark}}           & \color{Green}{\cmark}           \\
				\multicolumn{1}{l|}{CVE-2020-9548}  & \multicolumn{1}{c|}{\color{Green}{\cmark}}          & \multicolumn{1}{c|}{\color{Green}{\cmark}}          & \multicolumn{1}{c|}{\color{Green}{\cmark}}           & \color{Green}{\cmark}           \\
				\multicolumn{1}{l|}{CVE-2020-10650} & \multicolumn{1}{c|}{\color{Green}{\cmark}}          & \multicolumn{1}{c|}{\color{Green}{\cmark}}          & \multicolumn{1}{c|}{\color{Green}{\cmark}}           & \color{Green}{\cmark}          \\
				\multicolumn{1}{l|}{CVE-2020-10672} & \multicolumn{1}{c|}{\color{Green}{\cmark}}          & \multicolumn{1}{c|}{\color{Green}{\cmark}}          & \multicolumn{1}{c|}{\color{Green}{\cmark}}           & \color{Green}{\cmark}           \\
				\multicolumn{1}{l|}{CVE-2020-10968} & \multicolumn{1}{c|}{\color{Green}{\cmark}}          & \multicolumn{1}{c|}{\color{Green}{\cmark}}          & \multicolumn{1}{c|}{\color{Green}{\cmark}}           & \color{Green}{\cmark}           \\
				\multicolumn{1}{l|}{CVE-2020-10969} & \multicolumn{1}{c|}{\color{Green}{\cmark}}          & \multicolumn{1}{c|}{\color{Green}{\cmark}}          & \multicolumn{1}{c|}{\color{Green}{\cmark}}           & \color{Green}{\cmark}           \\
				\multicolumn{1}{l|}{CVE-2020-11112} & \multicolumn{1}{c|}{\color{Green}{\cmark}}          & \multicolumn{1}{c|}{\color{Green}{\cmark}}          & \multicolumn{1}{c|}{\color{Green}{\cmark}}           & \color{Green}{\cmark}          \\
				\multicolumn{1}{l|}{CVE-2020-11113} & \multicolumn{1}{c|}{\color{Green}{\cmark}}          & \multicolumn{1}{c|}{\color{Green}{\cmark}}          & \multicolumn{1}{c|}{\color{Green}{\cmark}}           & \color{Green}{\cmark}           \\
				\multicolumn{1}{l|}{CVE-2020-11619} & \multicolumn{1}{c|}{\color{Green}{\cmark}}          & \multicolumn{1}{c|}{\color{Green}{\cmark}}          & \multicolumn{1}{c|}{\color{Green}{\cmark}}           & \color{Green}{\cmark}           \\
				\multicolumn{1}{l|}{CVE-2020-11620} & \multicolumn{1}{c|}{\color{Green}{\cmark}}          & \multicolumn{1}{c|}{\color{Green}{\cmark}}          & \multicolumn{1}{c|}{\color{Green}{\cmark}}           & \color{Green}{\cmark}          \\
				\multicolumn{1}{l|}{CVE-2020-14060} & \multicolumn{1}{c|}{-}           & \multicolumn{1}{c|}{\color{Green}{\cmark}}          & \multicolumn{1}{c|}{\color{Green}{\cmark}}           & \color{Green}{\cmark}           \\
				\multicolumn{1}{l|}{CVE-2020-14061} & \multicolumn{1}{c|}{-}           & \multicolumn{1}{c|}{\color{Green}{\cmark}}          & \multicolumn{1}{c|}{\color{Green}{\cmark}}           & \color{Green}{\cmark}           \\
				\multicolumn{1}{l|}{CVE-2020-14062} & \multicolumn{1}{c|}{-}           & \multicolumn{1}{c|}{\color{Green}{\cmark}}          & \multicolumn{1}{c|}{\color{Green}{\cmark}}           & \color{Green}{\cmark}           \\
				\multicolumn{1}{l|}{CVE-2020-14195} & \multicolumn{1}{c|}{-}           & \multicolumn{1}{c|}{\color{Green}{\cmark}}          & \multicolumn{1}{c|}{\color{Green}{\cmark}}           & \color{Green}{\cmark}           \\
				\multicolumn{1}{l|}{CVE-2020-24616} & \multicolumn{1}{c|}{\color{Green}{\cmark}}          & \multicolumn{1}{c|}{\color{Green}{\cmark}}          & \multicolumn{1}{c|}{\color{Green}{\cmark}}           & \color{Green}{\cmark}          \\
				\multicolumn{1}{l|}{CVE-2020-24750} & \multicolumn{1}{c|}{\color{Green}{\cmark}}          & \multicolumn{1}{c|}{\color{Green}{\cmark}}          & \multicolumn{1}{c|}{\color{Green}{\cmark}}           & \color{Green}{\cmark}          \\
				\rowcolor{gray!25}
				\multicolumn{5}{l}{jetty-server-9.4.10.v20180503}                                                                                                           \\
				\multicolumn{1}{l|}{CVE-2019-10241} & \multicolumn{1}{c|}{\color{Green}{\cmark}}          & \multicolumn{1}{c|}{\color{Green}{\cmark}}          & \multicolumn{1}{c|}{\color{Green}{\cmark}}           & \color{Red}{\xmark}           \\
				\multicolumn{1}{l|}{CVE-2019-10247} & \multicolumn{1}{c|}{-}           & \multicolumn{1}{c|}{\color{Green}{\cmark}}          & \multicolumn{1}{c|}{\color{Green}{\cmark}}           & \color{Green}{\cmark}           \\
				\multicolumn{1}{l|}{CVE-2019-17632} & \multicolumn{1}{c|}{-}           & \multicolumn{1}{c|}{\color{Red}{\xmark}}          & \multicolumn{1}{c|}{\color{Red}{\xmark}}           & \color{Red}{\xmark}          \\
				\rowcolor{gray!25}
				\multicolumn{5}{l}{jetty-servlet-9.4.10.v20180503}                                                                                                          \\
				\multicolumn{1}{l|}{CVE-2019-10241} & \multicolumn{1}{c|}{\color{Green}{\cmark}}          & \multicolumn{1}{c|}{\color{Red}{\xmark}}          & \multicolumn{1}{c|}{\color{Red}{\xmark}}           & \color{Red}{\xmark}           \\
				\rowcolor{gray!25}
				\multicolumn{5}{l}{jetty-util-9.4.10.v20180503}                                                                                                             \\
				\multicolumn{1}{l|}{CVE-2019-10241} & \multicolumn{1}{c|}{-}           & \multicolumn{1}{c|}{\color{Red}{\xmark}}          & \multicolumn{1}{c|}{-\footnote{\label{nvdFootnote}The NVD entry did not contain any version indications}}            & \color{Red}{\xmark}           \\
				\multicolumn{1}{l|}{CVE-2019-10246} & \multicolumn{1}{c|}{-}           & \multicolumn{1}{c|}{\color{Red}{\xmark}}          & \multicolumn{1}{c|}{-\footref{nvdFootnote}}            & \color{Red}{\xmark}          \\
				\multicolumn{1}{l|}{CVE-2021-44832} & \multicolumn{1}{c|}{-}           & \multicolumn{1}{c|}{\color{Red}{\xmark}}          & \multicolumn{1}{c|}{\color{Red}{\xmark}}           & \color{Red}{\xmark}          \\
				\rowcolor{gray!25}
				\multicolumn{5}{l}{netty-all-4.0.36.Final}                                                                                                                  \\
				\multicolumn{1}{l|}{CVE-2021-21290} & \multicolumn{1}{c|}{-}           & \multicolumn{1}{c|}{\color{Green}{\cmark}}          & \multicolumn{1}{c|}{\color{Green}{\cmark}}           & \color{Green}{\cmark}           \\
				\rowcolor{gray!25}
				\multicolumn{5}{l}{poi-ooxml-3.14}                                                                                                                          \\
				\multicolumn{1}{l|}{CVE-2019-12415} & \multicolumn{1}{c|}{\color{Green}{\cmark}}          & \multicolumn{1}{c|}{\color{Green}{\cmark}}          & \multicolumn{1}{c|}{\color{Green}{\cmark}}           & \color{Green}{\cmark}          \\
				\rowcolor{gray!25}
				\multicolumn{5}{l}{undertow-core-1.4.23.Final}                                                                                                              \\
				\multicolumn{1}{l|}{CVE-2017-2670}  & \multicolumn{1}{c|}{-}           & \multicolumn{1}{c|}{\color{Red}{\xmark}}          & \multicolumn{1}{c|}{\color{Red}{\xmark}}           & \color{Green}{\cmark}          
			\end{tabular}
		}
		\label{tab:jaralyzerOnlyResults}
	\end{subtable}%
	\hfill
	\begin{subtable}{.495\linewidth}
		\caption{\textsc{Steady} reports}
		\resizebox{\linewidth}{!}{
			\begin{tabular}{lcccc}
				\multicolumn{1}{l|}{\textbf{CVE}}     & \multicolumn{1}{c|}{\textbf{KB}} & \multicolumn{1}{c|}{\textbf{GA}} & \multicolumn{1}{c|}{\textbf{NVD}} & \textbf{SV}           \\ \hline
				\rowcolor{gray!25}
				\multicolumn{5}{l}{bcprov-jdk15on-1.58}                                                                                                                                 \\
				\multicolumn{1}{l|}{CVE-2015-6644}    & \multicolumn{1}{c|}{-}           & \multicolumn{1}{c|}{\color{Red}{\xmark}}          & \multicolumn{1}{c|}{\color{Red}{\xmark}}           & -                     \\
				\multicolumn{1}{l|}{CVE-2016-1000338} & \multicolumn{1}{c|}{-}           & \multicolumn{1}{c|}{\color{Red}{\xmark}}          & \multicolumn{1}{c|}{\color{Red}{\xmark}}           & \color{Red}{\xmark}                     \\
				\multicolumn{1}{l|}{CVE-2016-1000339} & \multicolumn{1}{c|}{-}           & \multicolumn{1}{c|}{\color{Red}{\xmark}}          & \multicolumn{1}{c|}{\color{Red}{\xmark}}           & \color{Red}{\xmark}                     \\
				\multicolumn{1}{l|}{CVE-2016-1000340} & \multicolumn{1}{c|}{-}           & \multicolumn{1}{c|}{\color{Red}{\xmark}}          & \multicolumn{1}{c|}{\color{Red}{\xmark}}           & \color{Red}{\xmark}                     \\
				\multicolumn{1}{l|}{CVE-2016-1000341} & \multicolumn{1}{c|}{-}           & \multicolumn{1}{c|}{\color{Red}{\xmark}}          & \multicolumn{1}{c|}{\color{Red}{\xmark}}           & \color{Red}{\xmark}                     \\
				\multicolumn{1}{l|}{CVE-2016-1000342} & \multicolumn{1}{c|}{-}           & \multicolumn{1}{c|}{\color{Red}{\xmark}}          & \multicolumn{1}{c|}{\color{Red}{\xmark}}           & \color{Red}{\xmark}                     \\
				\multicolumn{1}{l|}{CVE-2016-1000343} & \multicolumn{1}{c|}{-}           & \multicolumn{1}{c|}{\color{Red}{\xmark}}          & \multicolumn{1}{c|}{\color{Red}{\xmark}}           & \color{Red}{\xmark}                     \\
				\multicolumn{1}{l|}{CVE-2016-1000344} & \multicolumn{1}{c|}{-}           & \multicolumn{1}{c|}{\color{Red}{\xmark}}          & \multicolumn{1}{c|}{\color{Red}{\xmark}}           & \color{Red}{\xmark}                     \\
				\multicolumn{1}{l|}{CVE-2016-1000345} & \multicolumn{1}{c|}{-}           & \multicolumn{1}{c|}{\color{Red}{\xmark}}          & \multicolumn{1}{c|}{\color{Red}{\xmark}}           & \color{Red}{\xmark}                     \\
				\multicolumn{1}{l|}{CVE-2016-1000346} & \multicolumn{1}{c|}{-}           & \multicolumn{1}{c|}{\color{Red}{\xmark}}          & \multicolumn{1}{c|}{\color{Red}{\xmark}}           & \color{Red}{\xmark}                     \\
				\multicolumn{1}{l|}{CVE-2016-1000352} & \multicolumn{1}{c|}{\color{Red}{\xmark}}          & \multicolumn{1}{c|}{\color{Red}{\xmark}}          & \multicolumn{1}{c|}{\color{Red}{\xmark}}           & \color{Red}{\xmark}                     \\
				\rowcolor{gray!25}
				\multicolumn{5}{l}{commons-compress-1.9}                                                                                                                                \\
				\multicolumn{1}{l|}{CVE-2019-12402}   & \multicolumn{1}{c|}{-}           & \multicolumn{1}{c|}{\color{Red}{\xmark}}          & \multicolumn{1}{c|}{\color{Red}{\xmark}}           & \color{Red}{\xmark}                    \\
				\rowcolor{gray!25}
				\multicolumn{5}{l}{groovy-all-2.4.7}                                                                                                                                    \\
				\multicolumn{1}{l|}{CVE-2015-3253}    & \multicolumn{1}{c|}{\color{Red}{\xmark}}          & \multicolumn{1}{c|}{\color{Red}{\xmark}}          & \multicolumn{1}{c|}{\color{Red}{\xmark}}           & \color{Red}{\xmark}                    \\
				\rowcolor{gray!25}
				\multicolumn{5}{l}{hibernate-validator-5.4.1.Final}                                                                                                                     \\
				\multicolumn{1}{l|}{CVE-2014-3558}    & \multicolumn{1}{c|}{\color{Red}{\xmark}}          & \multicolumn{1}{c|}{\color{Red}{\xmark}}          & \multicolumn{1}{c|}{\color{Red}{\xmark}}           & \color{Red}{\xmark}                     \\
				\rowcolor{gray!25}
				\multicolumn{5}{l}{httpclient-4.5.2}                                                                                                                                    \\
				\multicolumn{1}{l|}{CVE-2013-4366}    & \multicolumn{1}{c|}{\color{Red}{\xmark}}          & \multicolumn{1}{c|}{\color{Red}{\xmark}}          & \multicolumn{1}{c|}{\color{Red}{\xmark}}           & \color{Red}{\xmark}                    \\
				\rowcolor{gray!25}
				\multicolumn{5}{l}{jackson-databind-2.9.7}                                                                                                                              \\
				\multicolumn{1}{l|}{CVE-2017-17485}   & \multicolumn{1}{c|}{-}           & \multicolumn{1}{c|}{\color{Red}{\xmark}}          & \multicolumn{1}{c|}{\color{Red}{\xmark}}           & \color{Red}{\xmark}                     \\
				\multicolumn{1}{l|}{CVE-2018-5968}    & \multicolumn{1}{c|}{\color{Red}{\xmark}}          & \multicolumn{1}{c|}{\color{Red}{\xmark}}          & \multicolumn{1}{c|}{\color{Red}{\xmark}}           & \color{Red}{\xmark}                     \\
				\multicolumn{1}{l|}{CVE-2018-11307}   & \multicolumn{1}{c|}{\color{Red}{\xmark}}          & \multicolumn{1}{c|}{\color{Red}{\xmark}}          & \multicolumn{1}{c|}{\color{Red}{\xmark}}           & \color{Red}{\xmark}                     \\
				\multicolumn{1}{l|}{CVE-2018-12022}   & \multicolumn{1}{c|}{\color{Red}{\xmark}}          & \multicolumn{1}{c|}{\color{Red}{\xmark}}          & \multicolumn{1}{c|}{\color{Red}{\xmark}}           & \color{Red}{\xmark}                     \\
				\multicolumn{1}{l|}{CVE-2018-12023}   & \multicolumn{1}{c|}{\color{Red}{\xmark}}          & \multicolumn{1}{c|}{\color{Red}{\xmark}}          & \multicolumn{1}{c|}{\color{Red}{\xmark}}           & \color{Red}{\xmark}                     \\
				\rowcolor{gray!25}
				\multicolumn{5}{l}{jackson-dataformat-xml-2.9.3}                                                                                                                        \\
				\multicolumn{1}{l|}{CVE-2016-3720}    & \multicolumn{1}{c|}{-}           & \multicolumn{1}{c|}{\color{Red}{\xmark}}          & \multicolumn{1}{c|}{\color{Red}{\xmark}}           & \color{Red}{\xmark}                    \\
				\rowcolor{gray!25}
				\multicolumn{5}{l}{jetty-http-9.4.10.v20180503}                                                                                                                         \\
				\multicolumn{1}{l|}{CVE-2015-2080}    & \multicolumn{1}{c|}{\color{Red}{\xmark}}          & \multicolumn{1}{c|}{\color{Red}{\xmark}}          & \multicolumn{1}{c|}{\color{Red}{\xmark}}           & \color{Red}{\xmark}                    \\
				\multicolumn{1}{l|}{CVE-2017-7657}    & \multicolumn{1}{c|}{\color{Green}{\cmark}}          & \multicolumn{1}{c|}{\color{Red}{\xmark}}          & \multicolumn{1}{c|}{\color{Green}{\cmark}}           & \color{Red}{\xmark}                     \\
				\rowcolor{gray!25}
				\multicolumn{5}{l}{jetty-server-9.4.10.v20180503}                                                                                                                       \\
				\multicolumn{1}{l|}{CVE-2016-4800}    & \multicolumn{1}{c|}{-}           & \multicolumn{1}{c|}{\color{Red}{\xmark}}          & \multicolumn{1}{c|}{\color{Red}{\xmark}}           & \color{Red}{\xmark}                    \\
				\multicolumn{1}{l|}{CVE-2017-7656}    & \multicolumn{1}{c|}{\color{Green}{\cmark}}          & \multicolumn{1}{c|}{\color{Green}{\cmark}}          & \multicolumn{1}{c|}{\color{Green}{\cmark}}           & \color{Green}{\cmark}                     \\
				\multicolumn{1}{l|}{CVE-2017-7657}    & \multicolumn{1}{c|}{\color{Green}{\cmark}}          & \multicolumn{1}{c|}{\color{Red}{\xmark}}          & \multicolumn{1}{c|}{\color{Green}{\cmark}}           & \color{Red}{\xmark}                     \\
				\multicolumn{1}{l|}{CVE-2017-7658}    & \multicolumn{1}{c|}{\color{Green}{\cmark}}          & \multicolumn{1}{c|}{\color{Green}{\cmark}}          & \multicolumn{1}{c|}{\color{Green}{\cmark}}           & \color{Green}{\cmark}                     \\
				\multicolumn{1}{l|}{CVE-2018-12538}   & \multicolumn{1}{c|}{-}           & \multicolumn{1}{c|}{\color{Green}{\cmark}}          & \multicolumn{1}{c|}{\color{Red}{\xmark}}           & \color{Red}{\xmark}                    \\
				\rowcolor{gray!25}
				\multicolumn{5}{l}{jetty-util-9.4.10.v20180503}                                                                                                                         \\
				\multicolumn{1}{l|}{CVE-2016-4800}    & \multicolumn{1}{c|}{-}           & \multicolumn{1}{c|}{\color{Red}{\xmark}}          & \multicolumn{1}{c|}{\color{Red}{\xmark}}           & \color{Red}{\xmark}                    \\
				\multicolumn{1}{l|}{CVE-2017-9735}    & \multicolumn{1}{c|}{-}           & \multicolumn{1}{c|}{\color{Red}{\xmark}}          & \multicolumn{1}{c|}{\color{Red}{\xmark}}           & \color{Red}{\xmark}                     \\
				\multicolumn{1}{l|}{CVE-2018-12536}   & \multicolumn{1}{c|}{-}           & \multicolumn{1}{c|}{\color{Red}{\xmark}}          & \multicolumn{1}{c|}{\color{Red}{\xmark}}           & \color{Green}{\cmark}                     \\
				\rowcolor{gray!25}
				\multicolumn{5}{l}{okhttp-2.7.0}                                                                                                                                        \\
				\multicolumn{1}{l|}{CVE-2016-2402}    & \multicolumn{1}{c|}{\color{Green}{\cmark}}          & \multicolumn{1}{c|}{\color{Green}{\cmark}}          & \multicolumn{1}{c|}{\color{Green}{\cmark}}           & \color{Green}{\cmark}                    \\
				\rowcolor{gray!25}
				\multicolumn{5}{l}{spring-core-5.0.4.RELEASE}                                                                                                                           \\
				\multicolumn{1}{l|}{CVE-2015-0201}    & \multicolumn{1}{c|}{\color{Red}{\xmark}}          & \multicolumn{1}{c|}{\color{Red}{\xmark}}          & \multicolumn{1}{c|}{\color{Red}{\xmark}}           & \color{Red}{\xmark}                    \\
				\rowcolor{gray!25}
				\multicolumn{5}{l}{spring-oxm-5.0.4.RELEASE}                                                                                                                            \\
				\multicolumn{1}{l|}{CVE-2014-0054}    & \multicolumn{1}{c|}{\color{Red}{\xmark}}          & \multicolumn{1}{c|}{\color{Red}{\xmark}}          & \multicolumn{1}{c|}{\color{Red}{\xmark}}           & \color{Red}{\xmark}                    \\
				\multicolumn{1}{l|}{CVE-2014-0225}    & \multicolumn{1}{c|}{-}           & \multicolumn{1}{c|}{\color{Red}{\xmark}}          & \multicolumn{1}{c|}{\color{Red}{\xmark}}           & \color{Red}{\xmark}                    \\
				\multicolumn{1}{l|}{CVE-2014-3578}    & \multicolumn{1}{c|}{-}           & \multicolumn{1}{c|}{\color{Red}{\xmark}}          & \multicolumn{1}{c|}{\color{Red}{\xmark}}           & \color{Red}{\xmark}                    \\
				\rowcolor{gray!25}
				\multicolumn{5}{l}{spring-web-5.0.4.RELEASE}                                                                                                                            \\
				\multicolumn{1}{l|}{CVE-2013-6429}    & \multicolumn{1}{c|}{\color{Red}{\xmark}}          & \multicolumn{1}{c|}{\color{Red}{\xmark}}          & \multicolumn{1}{c|}{\color{Red}{\xmark}}           & \color{Red}{\xmark}                    \\
				\multicolumn{1}{l|}{CVE-2013-6430}    & \multicolumn{1}{c|}{-}           & \multicolumn{1}{c|}{\color{Red}{\xmark}}          & \multicolumn{1}{c|}{\color{Red}{\xmark}}           & \color{Red}{\xmark}                    \\
				\multicolumn{1}{l|}{CVE-2014-0054}    & \multicolumn{1}{c|}{-}           & \multicolumn{1}{c|}{\color{Red}{\xmark}}          & \multicolumn{1}{c|}{\color{Red}{\xmark}}           & \color{Red}{\xmark}                    \\
				\multicolumn{1}{l|}{CVE-2014-3578}    & \multicolumn{1}{c|}{-}           & \multicolumn{1}{c|}{\color{Red}{\xmark}}          & \multicolumn{1}{c|}{\color{Red}{\xmark}}           & \color{Red}{\xmark}                    \\
				\rowcolor{gray!25}
				\multicolumn{5}{l}{tomcat-embed-core-8.5.33}                                                                                                                            \\
				\multicolumn{1}{l|}{CVE-2020-13934}   & \multicolumn{1}{c|}{-}           & \multicolumn{1}{c|}{\color{Green}{\cmark}}          & \multicolumn{1}{c|}{\color{Green}{\cmark}}           & \color{Green}{\cmark} \\
				\rowcolor{gray!25}
				\multicolumn{5}{l}{undertow-core-1.4.23.Final}                                                                                                                          \\
				\multicolumn{1}{l|}{CVE-2018-14642}   & \multicolumn{1}{c|}{-}           & \multicolumn{1}{c|}{\color{Green}{\cmark}}          & \multicolumn{1}{c|}{-\footref{nvdFootnote}}            & \color{Green}{\cmark} \\
				\multicolumn{1}{l|}{CVE-2019-3888}    & \multicolumn{1}{c|}{-}           & \multicolumn{1}{c|}{\color{Green}{\cmark}}          & \multicolumn{1}{c|}{\color{Green}{\cmark}}           & \color{Green}{\cmark} \\
				\multicolumn{1}{l|}{CVE-2020-10705}   & \multicolumn{1}{c|}{-}           & \multicolumn{1}{c|}{\color{Green}{\cmark}}          & \multicolumn{1}{c|}{\color{Green}{\cmark}}           & \color{Green}{\cmark}
			\end{tabular}
			\label{tab:steadyOnlyReports}
		}
	\end{subtable} 
	\label{tab:unmodifiedResults}
\end{table}
\endgroup

Tables~\ref{tab:jaralyzerOnlyResults} and~\ref{tab:steadyOnlyReports} present the CVE entries only reported by one of the tools. 
There are 43 CVE entries, which \jaralyzer reports but \steady does not (Table~\ref{tab:jaralyzerOnlyResults}).
Conversely, \steady reports 44 CVE entries that \jaralyzer does not (Table~\ref{tab:steadyOnlyReports}).
As previously noted, we consider CVE entries that are part of both knowledge bases.

For each of the reported CVE entries we performed a manual investigation to determine whether it is a true or false report.
To do so we considered four different OSS vulnerability advisories.
In particular we considered the GitHub Advisory Database (GA)~\cite{githubAdvisory}, the National Vulnerability Database (NVD)~\cite{nvd}, the Snyk Vulnerability Database (SV)~\cite{snykAdvisory}, and \kb (KB)~\cite{ponta2019manually}.
Although we included the OSV Scanner in our experiments, we do not consider the OSV database in this experiment, as it primarily aggregates existing vulnerability data (e.g. NVD and GA) and does not provide independent classification of new vulnerabilities~\cite{osvAggregation}.
In fact, \kb contains not only fix commits that we used as described in Section~\ref{subsec:setup}, but also manual assessments indicating whether an artifact is affected by a specific CVE entry. 
\steady uses these manual assessments to enhance its detection performance and cope with its inability to compare source code with bytecode.
To not interfere with our experiments, we removed these manual assessments from \steady's and \jaralyzer's respective knowledge bases. 
However we use them for our manual investigation.
We consider a CVE report to be a true positive (false positive, resp.) for an artifact, if all advisories that contain the CVE entry agree on the artifact being affected (not affected, resp.) by the given CVE entry.
Two authors independently verified whether the advisories listed the respective CVE entry to be affecting the specific artifact and version as reported by either tool.
Furthermore, to analyze the causes of disagreement between the tools, we reviewed their individual scan reports and manually inspected the corresponding fix commits to identify potential sources of misclassification. 
Unlike \jaralyzer, \steady produces scan reports with limited details, which makes it difficult to pinpoint the exact reasons behind erroneous results. 
The reports only list the CVE entries assigned to the scanned artifacts and classify them as either ``vulnerable'' or ``unknown''. 
In 37 of 44 cases, \steady marked the entries as ``unknown'' and requested manual analysis.

Table~\ref{tab:jaralyzerOnlyResults} shows the CVE entries only reported by \jaralyzer, with corresponding artifacts in gray boxes.
According to our manual investigation, 35 out of 43 reports are true positives for \jaralyzer and false negatives for \steady.
Note that 30 of them affect jackson-databind. 
\steady misses those, even though source code is available, because it cannot account for code transformations outside of methods, such as field initializations or initializer blocks, due to its inability to directly compare bytecode with source code.
These jackson-databind vulnerability fixes consist of either changing a regular expression stored in a field or adding entries to serialization blocklists in static initializers.
Since \jaralyzer operates on bytecode rather than source code, and because the Java compiler inlines field initializations and static initializer code into methods, in contrast to \steady, it natively supports such cases.

However, we also uncovered five cases where \jaralyzer reported false positives, according to the advisories. 
All can be attributed to erroneously reporting a CVE for multiple modules of the same project, due to spurious fix commits that contain changes unrelated to the actual fix as described in Section~\ref{subsec:rqone}, or because of incorrect re-packaging matches due to the threshold configurations.
For CVE-2016-6793, which \jaralyzer reported as affecting commons-fileupload-1.3.2, we made an interesting observation.
While this CVE entry actually affects Apache Wicket\footnote{\url{https://security.snyk.io/vuln/SNYK-JAVA-ORGAPACHEWICKET-31022}}, \jaralyzer reports it because it detected a re-packaging of the class affected by the vulnerability. 
Further investigation revealed that the developers of Apache Wicket copied vulnerable code from commons-fileupload into the Wicket project.
The vulnerability within commons-fileupload received its own CVE entry\footnote{\url{https://security.snyk.io/vuln/SNYK-JAVA-COMMONSFILEUPLOAD-30401}}, even though they describe the same vulnerability.
The CVE entry for commons-fileupload is not part of \kb and thus not reported by either tool.
Therefore, we do not consider this as a clear false positive, but rather as an unclear case.
We identified three other unclear cases, where the four advisories reported conflicting information.

To summarize, of the 43 CVE entries reported only by \jaralyzer, we classified 35 as true positives, 4 as false positives and 4 as unclear.

Table~\ref{tab:steadyOnlyReports} contains the CVE entries only reported by \steady.
According to our investigation, 33 out of 44 are false positives.
Only in 7 cases \steady reported a true positive that \jaralyzer missed.
Furthermore, 4 reports are unclear, since the advisories reported conflicting information.
According to \steady's scan report, it was unable to automatically classify the false positive cases and instead deferred their evaluation to manual analysis.
This might be due to an inability to retrieve the corresponding source code, inaccuracies in the retrieved code or due to ambiguous fix commit matches~\cite{steadyDoc}.
Due to the limited details in the scan reports the exact cause cannot be determined.
A similar issue appears for the five \jaralyzer reports classified as true positives, which \steady missed, excluding the ones related to jackson-databind. 
The corresponding fix commits involve changes to method bodies and should be thus detectable by \steady.
However, the scan report does not provide enough details to determine the exact reason for the missed detections.

Using the same methodology, we also validated the 61 CVE reports where both tools agree.
According to the advisories, both tools reported 28 true positives and 16 false positives.
The 17 remaining cases are unclear due to advisory disagreements.

As noted in Section~\ref{subsec:rqone}, the \achilles benchmark includes a ground truth that we initially excluded from our experiments due to identified inaccuracies.
Based on the findings from this research question, we conducted an additional validation of the provided ground truth. 
Overall, it contains classifications only for 35 of the 148 CVE reports (23\%) generated by either \jaralyzer or \steady, covering 9 of the 87 cases (9\%) where the tools disagreed and 26 of the 61 cases (42\%) where they agreed.
We further examined the 10 out of 35 ground truth classifications that contradicted the findings of \jaralyzer.
This involved a manual review of the corresponding reports, including an analysis of advisory data, inspection of code changes in the relevant fix commits, and investigation of related issue tracker entries. 
Our analysis indicated that 9 out of the 10 disagreeing classifications were likely incorrect within the ground truth. 
We reported these 9 discrepancies, as well as the 52 cases absent from \achilles, where \jaralyzer and all considered advisories agreed, to the authors of \achilles, who subsequently incorporated our feedback into the benchmark\footnote{\url{https://github.com/secure-software-engineering/achilles-benchmark-depscanners/pull/}
	\href{https://github.com/secure-software-engineering/achilles-benchmark-depscanners/pull/44}{44} \& 
	\href{https://github.com/secure-software-engineering/achilles-benchmark-depscanners/pull/45}{45}}.

\llbox{
	\jaralyzer outperforms \steady on unmodified dependencies.
	It identifies 35 unique true positives while reporting only 4 false positives.
	In contrast, \steady detects only 7 unique true positives but produces 33 false positives.
}

%% file: sections/evaluation/rq3.tex
\subsection{RQ3: Runtime Evaluation}
\label{subsec:rqthree}

We investigated how much time each of the scanners in our experimental setup (see Section~\ref{subsec:setup}) required to perform a vulnerability scan on the unmodified benchmark.
On average, an industry-grade Java project contains 36 dependencies~\cite{dann2021identifying}.
In comparison, our benchmark incorporates moderately more dependencies than a typical industry-grade project.

Code-centric scanners, in contrast to metadata-based scanners, typically come with significant overhead, which naturally increases the scan time.
\jaralyzer, e.g., needs to extract methods from the scanned JAR file, normalize and generate code property graphs for them.
These are all complex tasks, which take a non-trivial amount of time.
In total, \jaralyzer takes 131 seconds to scan the 56 OSS dependencies in our benchmark.
As stated previously, we execute \jaralyzer in its default mode \textit{and} re-packaging detection mode.
Executing \jaralyzer's re-packaging mode is only required to detect type-4 modified dependencies.
Running only its default mode, \jaralyzer does not miss any vulnerabilities in the unmodified benchmark and only takes 63 seconds.
Meanwhile, \steady takes 1,387 seconds, requiring more than ten times the scanning time for the benchmark.
Due to the lightweight approach behind metadata-based scanners, they require lower runtimes.
Specifically, \owaspDC required 9 seconds, OSV Scanner required 5 seconds, and the commercial scanner required 30 seconds.
GitHub Dependabot performs the analysis automatically, when a commit containing dependency metadata is pushed to GitHub, where analysis results can then be obtained via a request to the GitHub API. 
In contrast to the other scanners, this analysis does not run locally, but on the GitHub servers.
We thus were unable to measure its precise runtime and do not include it here. 

\llbox{
	For projects of above-average size, \jaralyzer takes between 63 and 131 seconds to complete the scan.
	Overall, \jaralyzer is more than ten times faster than \steady.
}

%% file: sections/threats.tex
\section{Threats to Validity}
\label{sec:threats}

A few potential threats might impact the validity of our experiments with \jaralyzer.
We used the \achilles benchmark generator to create the individual benchmarks serving as a baseline for our evaluation.
\achilles only provides versions of OSS artifacts released by 2021, thus for many artifacts we did not use the latest releases.
Furthermore, while we did not use \achilles inaccurate ground truth, we still relied on \achilles correctly modifying the dependency inclusions.
Although, our manual investigation revealed these to be most likely correct.
The number of CVE entries detected by \jaralyzer within type 4 modifications depends on the configured threshold parameters.
Lowering these thresholds may enable \jaralyzer to detect more CVE entries in the type 4 benchmark but could also reduce its precision in the unmodified benchmark.
We selected the default values based on empirical testing with our benchmark.
However, this approach may lead to overfitting, meaning other datasets might achieve better performance with different values.
\jaralyzer was unable to compile 61 CVE entries from \kb, which we therefore excluded from RQ2. 
However, in cases where compilation fails or vulnerability fixes lack source code changes, the detection can be augmented by using metadata, as demonstrated by \steady.
Finally, the effectiveness of code-centric dependency scanning depends on the quality of fix commits.
When fix commits include spurious changes unrelated to the actual fix, the dependency scanning performance may be affected.

%% file: sections/relatedwork.tex
\section{Related Work}
\label{sec:relatedWork}
Current legislation like the EU's Cyber Resilience Act and the US's Cybersecurity Executive Order require software development companies to focus on the secure usage of OSS components.
Therefore many commercial and open-source tools have been developed to detect known-to-be-vulnerable OSS dependencies.
Within our study we used the popular metadata-based tools \owaspDC~\cite{owaspDC}, OSV Scanner~\cite{osvScanner} and GitHub Dependabot~\cite{dependabot}, as well as the code-centric tool \steady~\cite{ponta2020detection, ponta2018beyond, plate2015impact}.
Over the years, the term \textit{software composition analysis} (SCA) has emerged to describe dependency scanners, particularly in commercial contexts.
Popular commercial SCA tools include Endor Labs SCA~\cite{endorlabsScanner}, Snyk Open Source SCA~\cite{snykScanner}, Mend SCA~\cite{mendScanner} and Black Duck SCA~\cite{blackduckScanner}.

Some approaches under the name of patch presence testing attempt to decide whether a patch has been applied to a specific binary.
FIBER~\cite{zhang2018precise} generates signatures from the security patch's C/C++ source code and searches for its presence in the target binary.
Osprey~\cite{sun2021osprey} lifts the C/C++ binary into a platform-agnostic intermediate representation (IR) and performs the patch presence test based on this IR.
PDiff~\cite{jiang2020pdiff} performs patch presence testing within downstream kernel images by generating semantic summaries of the patch and checking whether the target image is closer to the image before or after the patch is applied.
$PS^3$~\cite{zhan2024ps3} uses semantics-level symbolic execution to extract signatures that are stable under different compiler options.
It then uses those signatures to test the presence of the patch within the target C/C++ binary.
BScout~\cite{dai2020bscout} uses feature-extraction and leverages line number information to match lines of Java source code, belonging to the patch, to lines of bytecode within the target file.
Doing so, Dai et al. can determine whether the patch has been applied or not.
PPT4J~\cite{pan2024ppt4j} employs a similar feature-extraction based approach, which extracts features that are stable through the compilation process and uses these features to determine whether the patch is applied or not.
FIBER, PDiff, and $PS^3$ target C/C++ binaries, BScout and PPT4J address Java applications.
In the future the above mentioned approaches could complimentary be integrated into \jaralyzer's vulnerability scanning stage to further aid with the CPG-based patch presence testing.
Since they do not perform the initial library matching nor address modifications, we did not include them into our evaluation, since a comparison with dependency scanners aiming for similar goals is more relevant and adequate.

Some approaches explicitly target third-party library (TPL) identification for Android to find licensing issues or vulnerable components usages.
OSSPolice~\cite{duan2017identifying} finds license violations and security risks within Android apps by maintaining a database of unique app features and comparing the target app to this database to determine the exact TPL used.
ATVHunter~\cite{zhan2021atvhunter} applies control-flow graph based feature-extraction to more reliably identify the correct TPL and version.
PHunter~\cite{xie2023precise} relies on special obfuscation-resilient features to determine whether patches are applied to an Android TPL, even with code obfuscations applied.

In recent years, researchers have published different approaches to identify security-relevant commits and resulting fix-commit databases.
Ponta et al.~\cite{ponta2019manually} created a manually-curated set of fix commits, called \kb, focusing on industry-relevant OSS projects.
Bhandari et al.~\cite{bhandari2021cvefixes} propose an approach that automatically extracts fix commits from CVE entries within the NVD.
They create the \textsc{CVEFixes} database, containing fix commits for 5,365 CVE entries affecting OSS projects of various programming languages.
Sabetta et al.~\cite{sabetta2024known} propose the rule-based tool \textsc{Prospector}, which tries to automatically find and rank fix commits for specified vulnerability identifiers.
Akhoundali et al.~\cite{akhoundali2024morefixes} propose an approach that uses \textsc{Prospector} and create the \textsc{MoreFixes} database, containing fix commits for 26,617 CVEs affecting a diverse set of OSS projects.

%% file: sections/conclusion.tex
\section{Conclusion}
\label{sec:conclusion}

This paper presents \jaralyzer, a novel \textit{bytecode-centric} dependency scanner for Java.
While there are many open-source and commercial dependency scanners available, none of the existing tools is able to effectively deal with modified dependencies.
\jaralyzer is able to largely overcome these challenges.
By adopting an approach independent of the dependency's metadata or source code availability and instead directly analyzing its bytecode, \jaralyzer effectively detects known-to-be-vulnerable dependencies in Java projects, even across all common types of modifications.

An evaluation performed on 56 popular OSS artifacts, including five state-of-the-art dependency scanners, showcased that, while all other tools suffer performance deterioration when dependencies are modified, \jaralyzer's performance remains stable.
In few cases, it was even able to exploit the modifications and report fewer false positives.
But even when dependencies are unmodified, \jaralyzer outperforms the state-of-the-art code-centric dependency scanner \steady, by detecting more vulnerabilities and producing fewer false alerts.
These results position \jaralyzer to be the code-centric dependency scanner of choice for Java projects, especially when it comes to identifying modified dependencies.